\documentclass[aps,prb,twocolumn,showpacs,epsfig,floats]{revtex4}
\usepackage{amsmath}
\usepackage{color}
\usepackage{graphicx}
\usepackage{amsfonts}
\usepackage{mathrsfs}
\usepackage{amsmath}

\begin{document}
\title{Quantum dynamics with stochastic reset}

\author{B. Mukherjee$^{1}$, K. Sengupta$^{1}$, and Satya N. Majumdar$^{2}$}

\affiliation { $^{1}$Theoretical Physics Department, Indian
Association for the Cultivation of Science, Jadavpur,
Kolkata-700032, India. \\
$^{2}$Univ. Paris-Sud, CNRS, LPTMS, UMR 8626, Orsay F-01405,
France.}

\date{\today}
\begin{abstract}

We study non-equilibrium dynamics of integrable and non-integrable
closed quantum systems whose unitary evolution is interrupted with
stochastic resets, characterized by a reset rate $r$, that project
the system to its initial state. We show that the steady state
density matrix of a non-integrable system, averaged over the reset
distribution, retains its off-diagonal elements for any finite $r$.
Consequently a generic observable $\hat O$, whose expectation value
receives contribution from these off-diagonal elements, never
thermalizes under such dynamics for any finite $r$. We demonstrate
this phenomenon by exact numerical studies of experimentally
realizable models of ultracold bosonic atoms in a tilted optical
lattice. For integrable Dirac-like fermionic models driven
periodically between such resets, the reset-averaged steady state is
found to be described by a family of generalized Gibbs ensembles
(GGE s) characterized by $r$. We also study the spread of particle
density of a non-interacting one-dimensional fermionic chain,
starting from an initial state where all fermions occupy the left
half of the sample, while the right half is empty. When driven by
resetting dynamics, the density profile approaches at long times to
a nonequilibrium stationary profile that we compute exactly. We
suggest concrete experiments that can possibly test our theory.

\end{abstract}

%\date{}
\maketitle
\section{Introduction}

Non-equilibrium dynamics of closed quantum systems has been a
subject of intense theoretical and experimental studies in recent
years [\onlinecite{rev1,rev2,rev3,rev4,rev5}]. The initial
theoretical endeavor in this direction focussed on the study of
ramps and quenches through quantum critical points and surfaces
[\onlinecite{subir0, pascal1, kibble1,zureck1,anatoly1,ks1,ks2,
ks3,degrandi1,sumit1}]. The former class of studies investigated the
excitation density and residual energies of a quantum system after a
ramp. In the presence of a quantum critical point or surface which
is traversed during the ramp, such quantities exhibit power law
variation with the ramp rate with universal exponents
[\onlinecite{rev1,kibble1,zureck1,anatoly1,ks1,ks2,
ks3,degrandi1,sumit1}]. The study of long-time behavior of quantum
systems after a sudden quench and the nature of the resultant steady
states (provided they exist) have been some of the central issues
addressed in the latter class of studies
[\onlinecite{subir0,pascal1,degrandi1, rev5}].

It is well-known that the nature of these steady states depends on
whether such systems are integrable. The dynamics of integrable
systems is typically non-ergodic due to the presence of large number
of quasi-local conserved quantities $Q_i$. The presence of such
conserved quantities implies that integrable systems, taken out of
equilibrium, relax to steady states whose precise form depend on
$Q_i$. The density matrix describing such steady states may be
expressed as $\rho \sim \exp[-\sum_i \lambda_i Q_i]$, where the
parameters $\lambda_i$ are determined from initial values of $Q_i$
[\onlinecite{rev1,rev5}]. Such a form of the steady state density
matrix follows from entropy maximization in the presence of the
conserved quantities $Q_i$. The corresponding ensemble is termed as
generalized Gibbs ensemble (GGE)[\onlinecite{rev1}].

In contrast, for non-integrable systems, one typically reaches a
thermal distribution at late times where the system is described by
a diagonal density matrix with an effective temperature
[\onlinecite{mrigol1,mrigol2,rev5}]. In these thermal steady states,
the expectation value of any typical local observable $\hat O$ of
the system is expected to agree with that obtained by averaging over
a microcannonical ensemble. The above-mentioned feature can be
viewed as a consequence of the eigenstate thermalization
hypothesis(ETH). This hypothesis follows from the fact that for a
generic quench protocol, the post-quench dynamics of any state with
fixed initial energy $\epsilon$ is governed by the final
Hamiltonian. Thus such dynamics preserves $\epsilon$. The system
under such dynamics explores all eigenstates in the vicinity of
$\epsilon$. Since such dynamics is ergodic over all eigenstates
within a narrow energy shell of $\epsilon$ and $\epsilon+\delta
\epsilon$, the time average of any observable can be equated to the
microcanonical ensemble average over these states: $\langle m|\hat
O|n\rangle \simeq O_{\rm mc}(\bar \epsilon) \delta_{mn}$. Here $\bar
\epsilon = (\epsilon_m + \epsilon_n)/2$ and $O_{\rm mc} = {\rm Tr}
[\rho_{\rm mc}(\bar \epsilon) \hat O]$ is the expectation value of
$\hat O$ as obtained from a microcannonical ensemble with energy
$\bar \epsilon$. ETH then states that the difference of $\langle
m|\hat O|n\rangle$ from $O_{\rm mc}(\bar \epsilon)$ must vanish in
the thermodynamic limit [\onlinecite{mrigol1,
mrigol2,srednicci1,rev5}].

More recently, the study of such long-time behavior for periodically
driven systems has also been undertaken [\onlinecite{rev1,arnabd1,
asen0}]. It is well-known that for non-integrable systems a periodic
drive heats up the system and takes it to an infinite temperature
fixed point. However, for integrable models this is not the case,
and periodically driven integrable systems may exhibit novel steady
states [\onlinecite{arnabd1}]. The behavior of such systems in the
presence of a stochastic aperiodicity superposed over a periodic
drive has also been studied recently [\onlinecite{asen0}].

In a different classical context, a number of recent studies have
found that a classical system evolving under its own natural
dynamics, when interrupted stochastically at random times following
which the system is {\it reset} to its initial condition, evolves at
long times into a nontrivial nonequilibrium stationary
state~[\onlinecite{EM12011,EM22011,EM32014,GMS2014,DHP2014,MSS2015,MSS22015,Pal2015,EM2016,NG2016,BEM2016,PKE2016,FBGM2017,MT2017,RG2017}].
This is most easily seen in the case of a single diffusive particle
on a line, starting at $x=0$. The position of a particle at time
$t$, without any resetting, has the standard probability density
$P(x,t)= e^{-x^2/{4Dt}}/\sqrt{4\pi D t}$ at time $t$, where $D$ is
the diffusion constant. If the particle is now reset to $x=0$ after
a random exponentially distributed time with rate $r$, the
probability density at long times becomes
time-independent~[\onlinecite{EM12011,EM22011}] and is given by:
$P_{\rm stat}(x)= (\alpha_0/2)\, \exp[-\alpha_0\,|x|]$, with
$\alpha_0=\sqrt{r/D}$. This result generalizes easily to higher
dimensions~[\onlinecite{EM32014}]. The approach to this stationary
state was shown to have an unusual relaxation dynamics, accompanied
by a dynamical phase transition~[\onlinecite{MSS2015}]. Such
resetting dynamics also has important consequences for search
processes: instead of searching for a target by pure diffusion, it
is more efficient to reset the searcher at its initial position at
random times---a lot of recent studies have demonstrated this in a
number of classical systems by studying the associated first-passage
problems~[\onlinecite{Rednerbook,BMSreview}] in the presence of
resetting~[\onlinecite{EM12011,EM22011,EMM2013,WEM2013,MV2013,KMSS2014,RUK2014,CS2015,R2016,BBS2016,MPV2017,PR2017}].
Functionals of Brownian motion with resetting have also been studied
recently~[\onlinecite{MST2015,HT2017,HMMT2018}]. Another interesting
observation is how resetting leads to novel stochastic
thermodynamics and the associated fluctuation theorems in classical
systems~[\onlinecite{FGS2016,PR22017}]. In addition, quantum systems
with projective measurements have been studied in the context of
fluctuation theorems and statistics of energy transfer between the
system and the measurement apparatus [\onlinecite{sto1}]. Moreover,
there have been several recent studies on quantum first detection
problems which involves interrupting unitary evolution of a quantum
system with projective measurements [\onlinecite{dhar1,eli1}].
However, the analog of a nonequilibrium stationary state induced by
random resettings is yet to be explored, to the best of our
knowledge, for closed quantum systems that undergo unitary evolution
in the absence of resetting.

In this work, we study the dynamics of integrable and non-integrable
quantum systems whose unitary evolution is interrupted by stochastic
resets, characterized by a reset rate $r$. We consider each reset to
project the wavefunction of the system to its initial value at
$t=0$. For a perfect reset protocol, which is what we shall be
mostly concerned with in this work, this is done with unit
probability. The main results obtained from our study of dynamics
with stochastic resets are as follows.

First, for non-integrable systems, we consider a generic initial
state which is not an eigenstate of the Hamiltonian controlling its
unitary evolution. We study the evolution of this state in the
presence of a stochastic reset characterized by rate $r$. We
demonstrate that such dynamics leads to a reset averaged steady
state density matrix (provided such a steady state exists for
unitary evolution without reset) which retains its off-diagonal
elements. Thus such systems are not described by a diagonal density
matrix in their steady states. Consequently, the expectation value
of a typical observable does not reduce to its thermal steady state
value under such dynamics. Our result reproduces the diagonal
density matrix for the steady state for $r=0$ which coincides with
known results for standard unitary evolution of a quantum system
[\onlinecite{rev5}]. In addition, it also leads to the quantum Zeno
effect for $ r \to \infty$ [\onlinecite{zenoref}]; in such a
situation, the initial state does not evolve and the density matrix
of the system is same as the initial density matrix at $t=0$.

Second, for studying the reset dynamics of integrable systems we
consider two distinct models. The first of these constitutes a
system of free fermions occupying the left half of a one dimensional
(1D) chain at $t=0$. These fermions evolve under a nearest-neighbor
hopping Hamiltonian [\onlinecite{antal1,antal2,eisler1}]. We show
that interruption of the unitary evolution of these fermions with
stochastic reset leads to non-trivial modification of their reset
averaged density, $n_m(r)$, where $m$ is the site index and $r$ is
the reset rate. We also find an exact scaling function for $n_m(r)$
and show that it reproduces the known behavior of $n_m$ for $r=0$
and $r \to \infty$. The second model involves a periodically driven
Dirac Hamiltonian in $d$-dimensions whose unitary evolution is
interrupted by a reset after a random integer number of periods. The
unitary dynamics of such a Hamiltonian is controlled by a periodic
drive characterized by a time period $T$. We show that the reset
averaged steady state of such driven systems correspond to GGEs
characterized by a reset rate $r$. We demonstrate this by computing
non-trivial correlation functions of the model. We find that the
steady state values of these correlation functions, averaged over
the reset probability, are smooth functions of $r$ which
demonstrates the $r$ dependence of the underlying GGEs.

Third, we carry out exact numerical studies of post-quench dynamics
of experimentally realizable models of ultracold bosonic atoms in a
tilted optical lattice in the presence of resets. The low-energy
physics of these bosons can be described in terms of dipoles (bound
pair of bosons and holes)[\onlinecite{subir1, greiner1,greiner2}].
We show that the excitation density of these dipoles $n_d$, the
dipole density-density correlation function $C$, and the half-chain
entanglement entropy $S$ of the boson chain, averaged over the reset
probability distribution, interpolates continually between their
values of reset free dynamics ($r=0$) and the quantum Zeno limit ($r
\to \infty$). We discuss experiments in context of this boson model
which can test our theory.

The plan of the rest of the work is as follows. In Sec.\
\ref{genform}, we present the general formalism for time evolution
with reset for generic non-integrable quantum systems and
demonstrate that the resultant steady state density matrix, averaged
over reset probability, retains its off-diagonal elements for any
finite $r$. This is followed by Sec.\ \ref{intmod} where we discuss
the dynamics of integrable models, namely, the 1D fermion chain and
the $d$-dimensional Dirac fermions, under such reset. In Sec.\
\ref{tbh}, we address the dynamics of the  Bose-Hubbard model in a
tilted optical lattice in the presence of stochastic reset. Finally,
we chart out experiments which test our theory, discuss our main
results, and conclude in Sec.\ \ref{diss}. Some applications of
quantum dynamics with stochastic resets to single particle quantum
mechanical systems are discussed in the Appendix A. Some other
details concerning the derivation of a scaling function are provided
in Appendix B.

\section{General Formalism}
\label{genform}

In this section, we consider a generic non-integrable quantum system
with unitary evolution interrupted by stochastic resets. In what
follows, we shall first consider the case when the reset takes the
system to its initial ground state with unit probability. We shall
briefly comment on the case of imperfect resets (where the state of
system may projected to some other states with a small but non-zero
probability) later in this section.

The time evolution of our system is defined precisely as follows.
Consider a quantum system, with a given Hamiltonian $H(t)$ (which
can in general be time-dependent), prepared initially at $t=0$ in
the state $|\psi(0)\rangle$. Now, the state $|\psi(t)\rangle$
evolves from $t$ to $t+dt$ as follows:
\begin{eqnarray}
|\psi(t+dt)\rangle =
\begin{cases} |\psi(0)\rangle, \quad\quad\quad\quad\quad\quad {\rm
with}\,\, {\rm
prob.}\,\, r\, dt \\
\nonumber\\ [1-i H(t)\, dt]\, |\psi(t)\rangle \,\,\,\, {\rm with}\,\,
{\rm prob.}\,\, 1-r\,dt
\end{cases}
\label{reset_dyn.0}
\end{eqnarray}
where we have set $\hbar=1$ for convenience. Here $r\ge 0$ denotes
the reset rate with which the system is projected back to the
initial state. Thus, in a small time interval $dt$, the system
either goes back to its initial state with probability $r\,dt$, or,
with the complementary probability $(1-r\, dt)$, it evolves
unitarily with its Hamiltonian $H(t)$. The density matrix ${\hat
\rho}(t)$ of the system at fixed time $t$, assuming it is in a pure
state, is then given by
\begin{equation}
{\hat \rho}(t)= |\psi(t)\rangle \langle \psi(t)|\, .
\label{dens_matrix.0}
\end{equation}
Note that for $r=0$, we have a purely unitary evolution
and $|\psi(t)\rangle$ is given by
\begin{equation}
|\psi(t)\rangle_{r=0}= U(0,t)\, |\psi(0)\rangle
\label{state_evol.1}
\end{equation}
where the unitary operator is $U$ is given by
\begin{eqnarray}
U(t_1, t_2) = T_{t}\, \exp[- i \int_{t_1}^{t_2} H(t') dt' ],
\label{evoleq1}
\end{eqnarray}
with $T_t$ denoting the time ordering. However for any $r>0$, the
dynamics is a mixture of stochastic and deterministic evolution and
the density matrix in Eq.\ (\ref{dens_matrix.0}) is stochastic in
the sense that it varies from one realization of the reset process
to another. Hence, the observed density matrix at time $t$ is
obtained by averaging over all possible reset histories
\begin{equation}
\rho(t)= E\left[{\hat \rho}(t)\right]
\label{dens_matrix.1}
\end{equation}
where $E$ denotes the classical expectation value over all
stochastic evolutions. Our goal is to investigate how a nonzero
$r$ modifies the time evolution of the quantum state, or
equivalently the associated density matrix in Eq.\
(\ref{dens_matrix.1}). A possible way to realize this mixture of
deterministic and stochastic dynamics in Eq.\ (\ref{reset_dyn.0}) in
a realistic system will be discussed later.

To compute the time evolution of the density matrix $\rho(t)$ in
Eq.\ (\ref{dens_matrix.1}) in the presence of a finite resetting
rate $r$, we first make the following observation. The resetting
protocol essentially induces a renewal process in the sense that
after each reset, the system again evolves unitarily from the same
initial state without having any memory of what happened before the
last reset. Hence, given the observation time $t$, what really
matters is how much time has elapsed since the last reset till time
$t$. Clearly, this time $\tau$, since the last reset, is a random
variable $\tau\in [0,t]$, whose probability density $p(\tau|t)$
(given a fixed observation time $t$) can be estimated as follows.
Imagine time running backwards from $t$ and consider the event that
there is no reset in the interval $[t-\tau, t]$ followed by a reset
in the small time interval $d\tau$. Now, since the resetting is a
Poisson process with rate $r$, the probability that there is no
reset in $[t-\tau, t]$ is simply $e^{-r \tau}$. The probability of a
reset in $d\tau$ is just $r\,d\tau$. Hence, taking the product, the
probability of this event is $r\, e^{-r \tau}\, d\tau$. Hence we get
\begin{equation}
p(\tau|t)\, d \tau= r\, e^{-r\tau}\, d\tau \quad\quad 0\le \tau <
t \label{reset_prob.0}
\end{equation}
Integrating, we get
\begin{eqnarray}
\int_0 ^{t} p(\tau|t) d \tau &=& 1 - e^{-r t} \label{probeq1}
\end{eqnarray}
which shows that the pdf $p(\tau|t)$ is not normalized to unity,
because the right hand side of Eq.\ (\ref{probeq1}) is just the
probability that there is at least one reset in $[0,t]$. There is
however the possibility of having no reset in $[0,t]$: the
probability for this event is simply $e^{-r t}$. Hence, the pdf
normalized to unity, given a fixed $t$, can be written as
\begin{equation}
p(\tau|t)= r\, e^{-r\, \tau} + e^{-r\, t}\, \delta(\tau-t)\quad\, 0\le \tau\le t\, .
\label{probeq2}
\end{equation}
It is easy to check that $\int_0^{t} p(\tau|t)\, d\tau=1$. The delta
function term in Eq.\ (\ref{probeq2}) effectively describes the
probability of the event of having no reset in $[0,t]$. Note that by
making the observation time $t$ large enough we can arbitrarily
reduce the probability of zero reset, and get rid of the last term
in Eq.\ (\ref{probeq2}).

Now let us consider the unitary evolution of the system, following
the last reset till the observation time $t$. The density
matrix of the system at $t$, given that $\tau$ is the time elapsed
since the last reset, is simply
\begin{eqnarray}
{\hat \rho}(\tau|t) &=& U(0,\tau)\, \rho_0\,  U^{\dagger}(0,\tau)
\label{condevol}
\end{eqnarray}
where $U$ is the unitary operator in Eq.\ (\ref{evoleq1}) and
$\rho_0$ is the density matrix immediately after the last reset.
However, for perfect reset, $\rho_0= |\psi(0)\rangle \langle
\psi(0)|$ is just the initial density matrix (since the system was
projected to the initial state at the last reset). Thus, using Eq.\
(\ref{probeq2}) one finds that the density matrix $\rho(t)$ in Eq.\
(\ref{dens_matrix.1}) (upon averaging over the random variable
$\tau$ associated with the reset process) is given by
\begin{eqnarray}
\rho(t) &=& \int_0^{t} r e^{-r \tau}\, \rho(\tau|t)\, d \tau \nonumber\\
&& + e^{-r t}\, U(0,t)\, \rho_0\, U^{\dagger}(0,t)
 \label{denmat1}
\end{eqnarray}
where the last term corresponds to the event that there is no reset
within $[0,t]$. Now, at long times $t$, this last term vanishes exponentially
and the density matrix $\rho(t)$ approaches a stationary value (as $t\to \infty)$
\begin{eqnarray}
\rho_{\rm stat}  &=& \int_0^{\infty} r e^{-r \tau}\, U(0,\tau)
\rho_0 U^{\dagger}(0,\tau) d \tau \label{denmat2}
\end{eqnarray}
where the subscript ${\rm `stat'}$ stands for `stationary'.

We note that there are two distinct ways one can interpret Eq.\
(\ref{denmat2}). The first constitutes looking at $\rho_{\rm stat}$
as an ensemble average. To see this, we consider a unitarily
evolving system without reset evolving from $t=0$ with the initial
density matrix $\rho_0$. Now, imagine $N_0$ copies of the system
along with $N_0$ observers. Each observer measures, for the first
time, the density matrix at a preferred time $\tau$ and records the
outcome. The time of measurement, $\tau$, varies from one observer
to another and is distributed as $p_r(\tau)= r e^{-r \tau}$. An
average of the outcome of such single one-time measurement for all
$N_0$ observers for large $N_0$ (which is the same as averaging over
$\tau$) yields $\rho_{\rm stat}$. Note that in this interpretation,
there is no reset and the observers do not track the evolution of
the system after the measurement. Thus this procedure leads to
$\rho_{\rm stat}$ via ensemble averaging over $N_0$ copies of the
system in the limit of large $N_0$.

The second way to interpret Eq.\ (\ref{denmat2}) is as follows. We
consider a single copy of unitarily evolving system without reset.
The observer makes the first measurement at a random time $\tau$
(exponentially distributed with $p_r(\tau)$) and immediately after
the measurement resets it to the initial state. It is to be noted
that here measurement and reset constitute two separate processes;
the reset protocol is to be designed to project the state of the
system after the measurement to its initial state. This processes is
repeated for several times followed by an average over all
measurement data. This again leads to $\rho_{\rm stat}$ via time
averaging over measurements carried out on a single copy of the
systems. Thus the time and the ensemble averages are clearly
equivalent; both of them may be used to obtain Eq.\ (\ref{denmat2})
provided $\tau$ is chosen from the same exponential distribution.
This equivalence owes its origin to the fact that the time evolution
of the system between any two reset events is independent of any
other such evolution; hence these evolutions lead to a statistical
ensemble.

To investigate further the consequence of a nonzero $r$ in the
evolution of the density matrix, we consider the steady state
density matrix of generic non-integrable system (reached at long
times and provided that such a steady state exists) following a
quench in the absence of resets ($r=0$). It is well-known that such
steady state density matrices retain only diagonal terms (in the
eigenbasis of the Hamiltonian $H$ controlling the post-quench
evolution); all off-diagonal terms vanish. To see this, let us
consider an arbitrary initial quantum state $|\psi(0)\rangle$ given
by
\begin{eqnarray}
|\psi(0) \rangle &=& \sum_{\alpha} c_{\alpha} |\alpha\rangle,  \quad
H |\alpha \rangle = \epsilon_{\alpha} |\alpha \rangle  \label{wav1}
\end{eqnarray}
where $c_{\alpha} = \langle \alpha |\psi\rangle$ denotes the overlap
of the wavefunction with the eigenstate $| \alpha \rangle$ and
$\epsilon_{\alpha}$ is the corresponding eigenvalue. The elements of
the density matrix at any time $t$ in the energy eigenbasis under
such Hamiltonian evolution is thus given by
\begin{eqnarray}
\rho_{\alpha \beta}(t) = \langle \beta|\rho|\alpha\rangle &=& c_{\alpha}^{\ast} c_{\beta} e^{-i
\omega_{\beta \alpha} t} \label{dm1}
\end{eqnarray}
and $\omega_{\beta \alpha} = (\epsilon_{\beta} -
\epsilon_{\alpha})$. Now consider the fate of this matrix element at
long time by calculating a time average of $\rho_{\alpha \beta}$
\begin{eqnarray}
{\overline \rho_{\alpha \beta}}(t) &=& \lim_{T \to \infty} \frac{1}{T} \int_0^T dt \rho_{\alpha \beta}(t)  \nonumber\\
&=& |c_\alpha|^2 \delta_{\alpha \beta}   \label{sdm1}
\end{eqnarray}
We note that only the diagonal terms survive at long times and such
a density matrix typically signifies that the system at long times,
in its steady state, is described by a diagonal density matrix
$\rho_D = \rho_{\alpha \alpha} \delta_{\alpha \beta}$. The evolution
of a quantum statistical system to such a steady state essentially
signifies loss of phase information of the initial state. Thus in
the steady state, any quantum operator $\hat O$ of the system has an
expectation value (Eq.\ \ref{sdm1})
\begin{eqnarray}
\langle O\rangle &=& \sum_{\alpha \beta} O_{\alpha \beta}
\rho_{\beta \alpha} = \sum_{\alpha} |c_{\alpha}|^2 O_{\alpha \alpha}
= O_D, \label{opdiag}
\end{eqnarray}

Next let us consider the fate of such a density matrix in the
presence of a stochastic reset with $r>0$. Using Eqs.\ (\ref{dm1})
and (\ref{denmat2}), one finds that
\begin{eqnarray}
{\overline \rho_{\alpha \beta}} &=& \int_0^{\infty} d \tau r e^{- (r+ i \omega_{\beta \alpha}) \tau} c_{\alpha}^{\ast} c_{\beta} \nonumber\\
&=& (\rho_0)_{\alpha \beta} \frac{r}{r+ i \omega_{\beta \alpha}}  \quad  {\rm for} \, \, \beta \ne \alpha \nonumber\\
&=& \rho_D  = (\rho_0)_{\alpha \alpha}   \quad  {\rm for}\,\,
\alpha=\beta \label{dmsto1}
\end{eqnarray}
where the initial density matrix elements are given by
$(\rho_0)_{\alpha \beta}= c_{\alpha}^{\ast} c_{\beta}$. Thus we find
that the reset averaged steady state density matrix retains
off-diagonal elements under time evolution and is not diagonal in
the energy basis. We therefore conclude that stochastic resets leads
to novel steady state density matrices. Note that for $r \to 0$,
which signifies, on the average, a very long reset time, the
off-diagonal terms vanish and the density matrix recovers its
diagonal form as expected. In contrast, for $ r \to \infty$,  $
\rho_{\alpha \beta} \to (\rho_0)_{\alpha \beta}$. This is a
manifestation of quantum Zeno effect signifying a total freezing out
of the system dynamics for successive projections with very short
intermediate unitary evolution.

Before closing this section we note that the presence of such
off-diagonal terms in the steady state density matrix of the system
will show up in the expectation value of any generic operator of
such a quantum system. Using Eq.\ (\ref{opdiag}), it is easy to see
that
\begin{eqnarray}
\langle O \rangle &=& O_D + \sum_{\alpha \ne \beta}
c_{\alpha}^{\ast} c_{\beta} O_{\beta \alpha} \frac{r}{r+ i
\omega_{\beta \alpha} } \label{opexp}
\end{eqnarray}
We note that all operator expectations deviate from their diagonal
ensemble values signifying the presence of a steady state
characterized by a non-diagonal density matrix. Furthermore, for
operators which obey $O_{\alpha \beta}=O_{\beta \alpha}$, Eq.\
(\ref{opexp}) can be cast to a slightly more suggestive form
\begin{eqnarray}
\langle O \rangle &=& O_D + \sum_{\alpha \beta}
|c_{\beta}||c_{\alpha}| O_{\alpha \beta}
\frac{r}{r^2+ \omega_{\alpha \beta}^2} \nonumber\\
&& \times [r \cos(\theta_{\alpha \beta}) + \omega_{\alpha \beta}
\sin(\theta_{\alpha \beta}) ] \label{opexp2}
\end{eqnarray}
where $c_{\alpha}= |c_{\alpha}| \exp[i \theta_{\alpha}]$, and
$\theta_{\alpha \beta}=\theta_{\alpha}-\theta_{\beta}$. Thus for any
finite $r$, $\langle O \rangle \ne {\rm Tr} [\rho_{\rm D} {\hat O}]
= O_D $ and it receives contribution from the off-diagonal elements
of $\rho$. Thus a generic observable does not thermalize under such
dynamics.

For perfect resets that we have considered so far, the overlap
coefficient $c_{\alpha}$, for any $\alpha$, is determined completely by
the initial wavefunction of the system. In contrast, if the reset is
imperfect, $c_{\alpha}$ would be stochastic and one would
need to average over them with respect to some probability
distribution. In the extreme case when the reset projects the system to a
completely random state in the Hilbert space, such an average leads
to $ \langle O \rangle = \sum_{\alpha} O_{\alpha \alpha} \langle
|c_{\alpha}|^2\rangle$ since random phase fluctuations (fluctuations
in $\theta_{\alpha}$) would cancel the contribution of the
off-diagonal terms to $\langle O \rangle$. In this case, one would only
get diagonal contributions. However, a generic imperfect reset is
not this extreme case. In a generic case,
the system is projected to a state that is not a fully random state, and
the distribution of $c_{\alpha}$ is expected to
peak around
their initial values with a finite width. In that case, an average
over values of $c_{\alpha}$ will retain a finite off-diagonal
contribution of $\langle O \rangle$. Thus we expect the off-diagonal
elements of $\rho$ and their contributions to $\langle O \rangle$ to
be robust against moderate imperfection in the reset protocol. For
the rest of this work, we shall analyze the case of perfect resets.

\section{Integrable models}
\label{intmod}

In this section, we consider two integrable models. The first one
would constitute a chain of 1D fermions on a lattice while the
second would be the free spinless fermions obeying a Dirac-like
equation in $d$-dimensions.

\subsection{Fermion chain in 1D}
\label{fermchain}

\begin{figure}
\includegraphics[width=\linewidth]{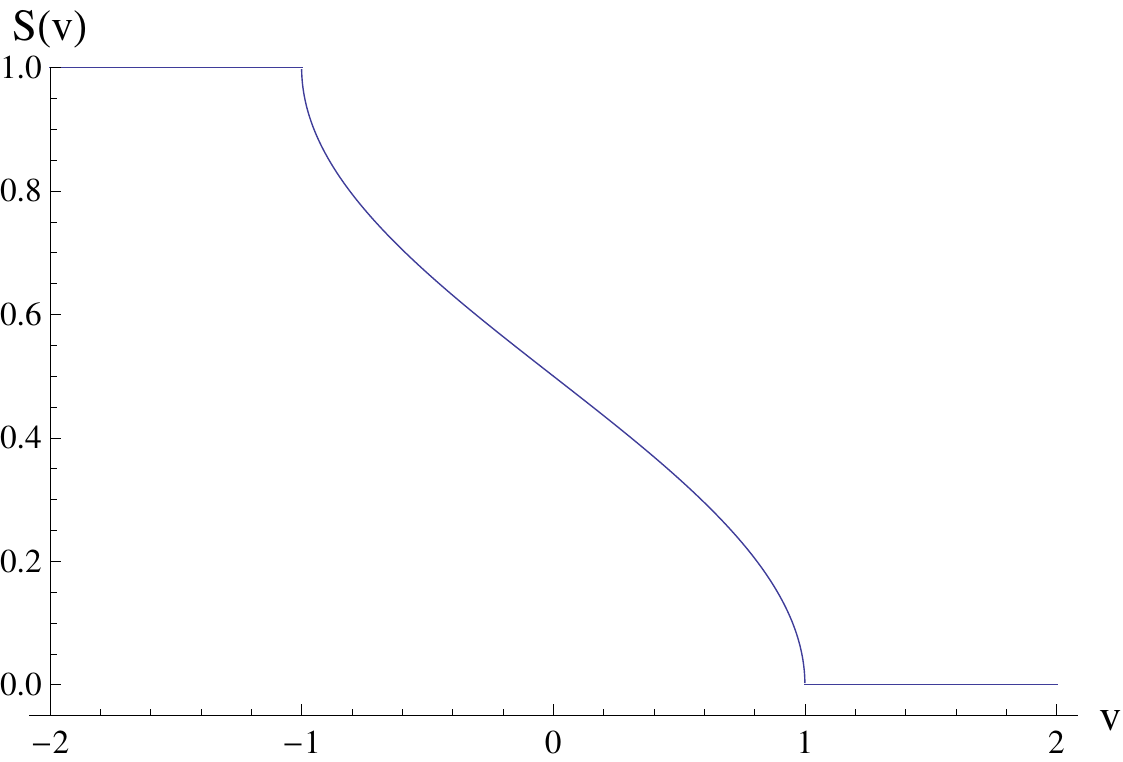}
\caption{The density profile at late times $t$ converges to a
scaling function $n_m(t)\to S\left(\frac{m}{t}\right)$. The figure
shows a plot of $S(v)$ vs. $v$.} \label{front_shape.fig}
\end{figure}

The fermion chain model that we consider consists of free spinless
fermions with nearest neighbor hopping such that
[\onlinecite{eisler1}]
\begin{eqnarray}
H= -(1/2) \sum_m (c_m^{\dagger} c_{m+1} +{\rm
h.c.})\label{chainham1}
\end{eqnarray}
where $c_m$ denotes the annihilation operator for a fermion on the
$m^{\rm th}$ site and the hopping amplitude of fermions is set to
$1/2$. The initial state for these fermions is chosen to be a step
function: each negative side (and the origin $0$) is occupied by a
fermion, while each positive site is empty, i.e., $\langle
c_{m}^{\dagger} c_n \rangle= \delta_{mn} \theta(-n)$, where $\theta$
denotes Heaviside step
function~[\onlinecite{antal1,antal2,eisler1}]. Thus, initially, the
density per site, on an average is $1/2$ and the subsequent unitary
evolution preserves the total number of particles. Using Fourier
transform, ${\tilde c}(k)= \sum_{m=-\infty}^{\infty} c_m\, e^{i\,
k\, m}$, one can easily diagonalize the Hamiltonian $H$. Also, using
this Fourier basis, one can easily express the fermion creation
operator at any time $t$ to be
\begin{eqnarray}
c_m(t) &=& e^{i H t} c_m(0) e^{-i H t} = \sum_n i^{n-m} J_{n-m}(t)
c_n(0), \nonumber\\
\end{eqnarray}
where $J_k(t)$ denotes Bessel function. Thus the expected density
$n_m(t)= \langle 0|c_{m}^{\dagger}(t)\, c_m(t)|0\rangle$ of the
fermions at site $m>0$ at any time $t$, under a Hamiltonian
evolution, can be expressed as~[\onlinecite{eisler1}]
\begin{equation}
n_m(t) =  \sum_{k=m}^{\infty} J_k^2(t)\, .
\label{fchaineq1}
\end{equation}
%\left( 1- J_0(t)^2\right) -\sum_{k=1}^{m-1} J_k^2(t)\quad {\rm for}\,\,m>0 \, .
For $m<0$, the density is simply
\begin{equation}
n_m(t)= 1- n_{1-m}(t)\quad {\rm for}\,\,m>0 \, .
\label{dens_negative}
\end{equation}
The average density profile $n_m(t)$ evolves with time. As time
$t\to \infty$, the average density $n_m(t)\to 1/2$ for every $m$,
i.e., the density profile becomes asymptotically flat with value
$1/2$ (since the evolution preserves the total number of particles).
However, at any finite time $t$, the density profile is rather
nontrivial. At any given time $t$, the density approaches
asymptotically to $1$ as $m\to -\infty$, while it vanishes
asymptotically as $m\to \infty$. However, away from these two
boundaries, inside the bulk, the density is different from $1$ or
$0$. This bulk region spreads around $m=0$ ballistically with time
$t$. Indeed, for large $t$ and large $|m|$, but with the ratio
$m/t=v$ fixed, by analyzing the asymptotics of Bessel functions in
Eqs.\ (\ref{fchaineq1}) and (\ref{dens_negative}), the density
profile $n_m(t)$ converges to a scaling form
\begin{equation}
n_m(t) \to S\left(\frac{m}{t}\right)
\label{dens_profile.1}
\end{equation}
where the scaling function $S(v)$, describing the shape of the bulk,
has a nontrivial form~[\onlinecite{antal1}]. For $v>0$,
\begin{eqnarray}
S(v) & = & \frac{1}{\pi}\, \cos^{-1}(v) \quad {\rm for}\,\, 0<v<1 \nonumber \\
&=& 0 \quad\quad\quad\quad\quad\,\, {\rm for}\, \, v\ge 1\,
\label{shape_positive}
\end{eqnarray}
while for $v<0$,
\begin{equation}
S(v)= 1-S(-v) \quad {\rm for}\,\, v<0\, .
\label{shape_negative}
\end{equation}
A plot of this shape scaling function $S(v)$ vs. $v$ is given in
Fig.\ \ref{front_shape.fig}. Thus, the density profile $n_m(t)$ for
late times $t$, has a nontrivial profile for $-t<m<t$, described by
the scaling function $S(v)$. The width of this bulk region increases
linearly with $t$ at late times $t$.

\begin{figure}
\includegraphics[width=\linewidth]{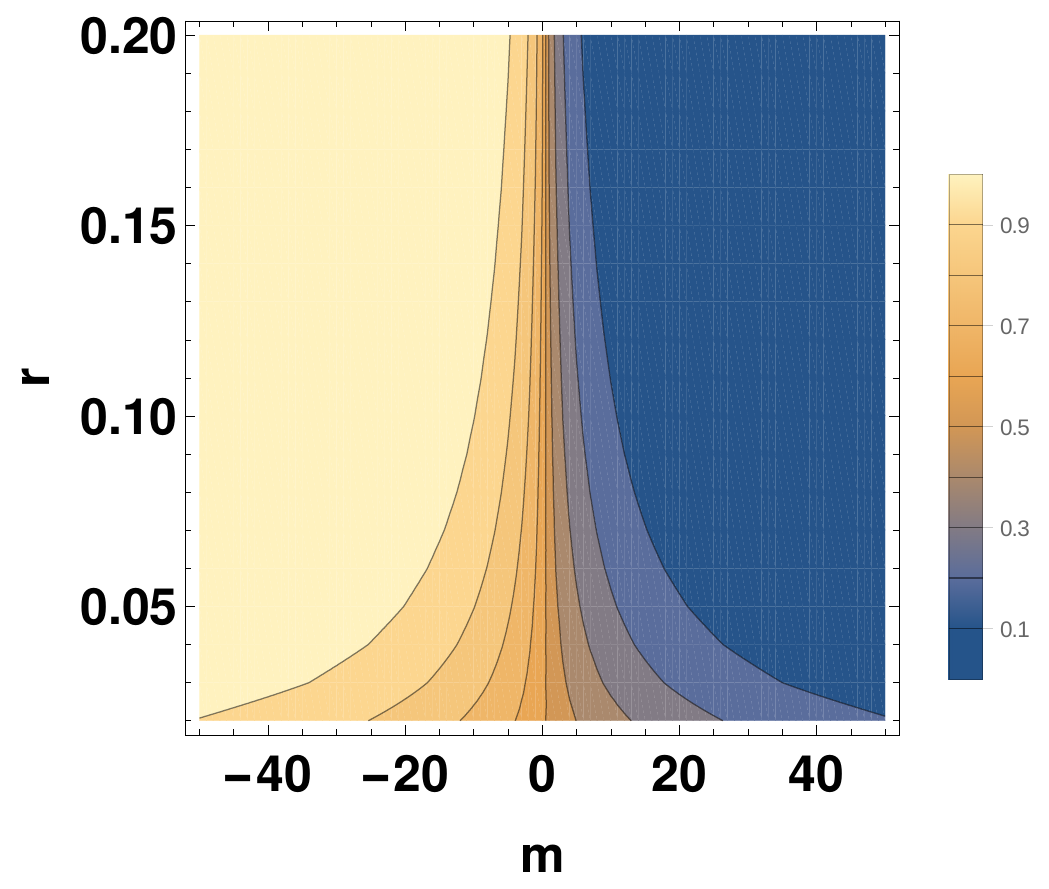}
%\rotatebox{0}{
%\includegraphics*[bb=0 0 246 238, width=\linewidth]{Fignew.jpg}}
\caption{Plot of $n_m(r)$ as a function of $r$ and $m$ in the $(m-r)$ plane, showing
exponential decay of $n_m(r)$ for large $m$ and large $r$. }
\label{fig1}
\end{figure}

We now consider switching on the reset mechanism with rate $r$,
discussed in the previous section. This resetting protocol will
drive the density to a nontrivial stationary state. To see this, we
compute the average density profile $n_m(r)$ of this fermionic chain
in presence of a finite reset rate $r$. The calculation is quite
easy and straightforward, given the general formalism in the
previous section. We start from the expression of the stationary
density matrix operator $\rho_{\rm stat}$ given in Eq.\
(\ref{denmat2}). The average stationary density profile, upon
choosing the resetting time distribution $p(\tau|t)$ as in Eq.\
(\ref{probeq2}) with $t\to \infty$ is then given by
\begin{eqnarray}
n_m(r) & = & \langle m| \rho_{\rm stat} |m\rangle = r
\int_0^{\infty} d\tau\, e^{-r \tau}\, n_m(\tau) \label{den_reset.1}
\end{eqnarray}
where $n_m(\tau)$ is the average density profile at time $\tau$
without reset, and is given explicitly in Eqs.\ (\ref{fchaineq1})
and (\ref{dens_negative}). Using this, we obtain, for $m>0$,
\begin{eqnarray}
n_m(r) =\sum_{k=m}^{\infty} C_k(r)
\label{fchaineq1.5}
\end{eqnarray}
where $C_k(r)= r\, \int_0^\infty J_k^2(\tau)\, e^{-r\tau}\, d\tau$.
For $m\le 0$, we have
\begin{equation}
n_{1-m}(r)= 1- n_m(r) \label{dens_negative_reset}
\end{equation}
which simply follows from Eq.\ (\ref{den_reset.1}) and the relation
in Eq.\ (\ref{dens_negative}). The function $C_k(r)$ can be
explicitly expressed as
\begin{eqnarray}
C_k(r)&= &\frac{4^m}{\pi} \Gamma^2(k+1/2) r^{-(2k)} \nonumber\\
&& \times _2 F_{1}(\frac{1}{2}+k, \frac{1}{2}+k, 1+2k;
-\frac{4}{r^2})
\label{fchaineq2}
\end{eqnarray}
where $\Gamma$ denotes the Gamma function and $_{2}F_1$ is the
regularized hypergeometric function. To get a feeling how the
spatial density profile looks, we provide, in Fig.\ \ref{fig1}, a
color plot of $n_m(r)$ as a function of $m$ and $r$ in the $(m-r)$
plane. The figure shows an exponential decay of $n_m(r)$ for large
$m$, as well as for large $r$. We note that for large $r$, $n_m(r)
\to 0$ for all $m$ which is the Zeno result showing that $n_m \to
n_m(t=0) = \theta(-m)$ in this limit. For $r \to 0$, $n_m(r)$
approaches the value $n_m(t)$ at large $t$ (without reset), namely
$n_m(r)\to 1/2$.

\begin{figure}
\includegraphics[width=\linewidth]{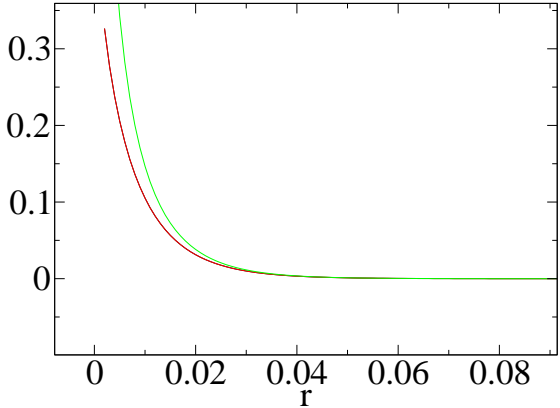}
%\rotatebox{0}{
%\includegraphics*[bb=0 0 246 238, width=\linewidth]{Fignew.jpg}}
\caption{A comparison of plot of $n_m(r)$ as a function of $r$ with
the analytic scaling form at large $m=100$. } \label{fig2}
\end{figure}

Obtaining precisely the large $m$ asymptotic behavior of $n_m(r)$,
for any fixed $r$, from the exact summation formula in Eqs.\
(\ref{fchaineq1.5}) and (\ref{fchaineq2}) turns out to be rather
cumbersome. However, in the limit of small $r$, the large $m$
behavior can be derived precisely as follows. For small $r$, the
integral in Eq.\ (\ref{den_reset.1}) is dominated by the large $t$
behavior of $n_m(t)$. Now, for large $m$ and large $t$, we can
replace $n_m(t)$ by its scaling form, $n_m(t)= S(m/t)$, where $S(v)$
is given in Eq.\ (\ref{shape_positive}). This gives, for large $m$
and $r\to 0$ (but fixed)
\begin{equation}
n_m(r) \approx  \frac{r}{\pi} \int_m^{\infty} d\tau\, e^{-r \tau}\, \cos^{-1}\left(\frac{m}{\tau}\right)\, .
\label{large_m.1}
\end{equation}
An integration by parts yields
\begin{equation}
n_m(r)= \frac{m}{\pi}
\int_m^{\infty} d\tau\, \frac{e^{-r\tau}}{\tau \sqrt{\tau^2-m^2}}\, .
\label{fchainint1}
\end{equation}
Next, we shift $\tau=m+x$, expand the integrand for large $m$ and carry out the
integration over the first few terms of the expansion.
This yields
\begin{equation}
n_m(r) \simeq   \frac{e^{-rm}}{\sqrt{2\pi r m}}\, \left[ 1- \frac{5}{8}\,\frac{1}{rm}
+ \frac{128}{129}\,\frac{1}{(rm)^2}+ \cdots \right]\, .
\label{fchainint2}
\end{equation}
Thus the leading behavior of the steady state density profile at
large $m$ is $n_m(r) \simeq e^{-mr}/\sqrt{2 \pi r m}$. This agrees
well with the exponential decay observed in numerics for large $m$
as long as $mr >1$, as can be seen in Fig.\ \ref{fig2}. Note that
the steady state is parametrized by the value of reset rate $r$.

\begin{figure}
\includegraphics[width=\linewidth]{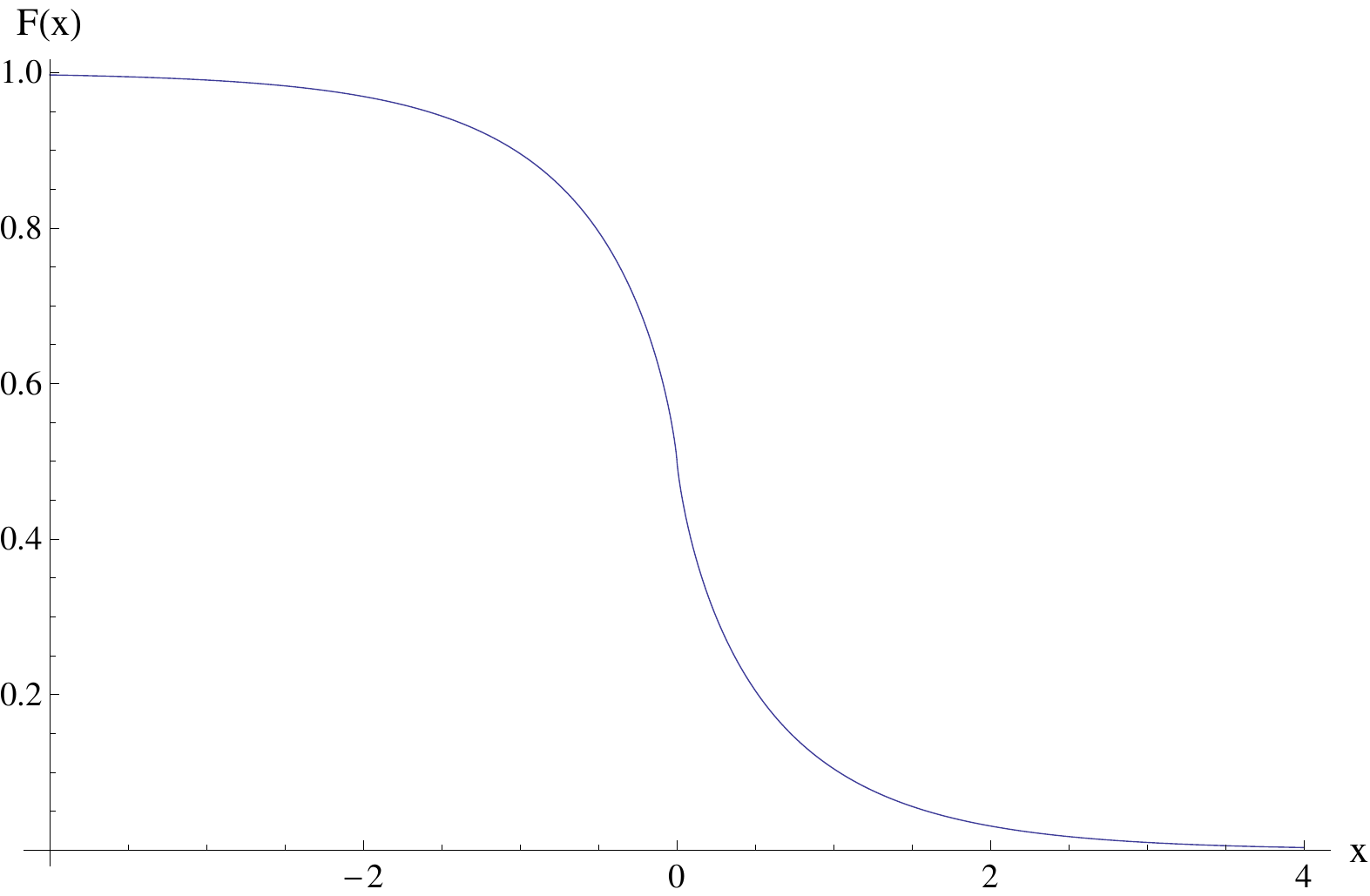}
\caption{The scaling function $F(x)$ vs. $x$ in Eq.
(\ref{scaling_function.2}). It satisfies the symmetry property:
$F(x)=1-F(-x)$.} \label{scaling_function.fig}
\end{figure}

The asymptotic result in Eq.\ (\ref{fchainint2}) for large $m$ and
fixed small $r$ suggests that there is a scaling limit $r\to 0$,
$m\to \infty$, but with the product $x=rm$ fixed such that the
density profile has a scale invariant form
\begin{equation}
n_m(r) \to F(r\,m)\, .
\label{scaling_function.1}
\end{equation}
Indeed, we find that this is the case with the full scaling function
$F(x)$ for all $x$, given explicitly by
\begin{equation}
F(x)= \frac{1}{\pi}\, \int_x^{\infty} K_0(|y|)\, dy\, .
\label{scaling_function.2}
\end{equation}
where $K_0(y)$ is the modified Bessel function of index $0$. A
derivation of this result in given in Appendix \ref{Fx_appendix}.
Note that $F(x)$ satisfies the symmetry relation,
\begin{equation}
F(-x)=1-F(x)
\label{scaling_symmetry}\, .
\end{equation}
A plot of this function is given in Fig.\
\ref{scaling_function.fig}. Using the known asymptotics of $K_0(y)$,
one can easily show that as $x\to \infty$
\begin{equation}
F(x) \to \frac{e^{-x}}{\sqrt{2\pi x}}\, \left[1- \frac{5}{8}\frac{1}{x}
+ \frac{128}{129}\frac{1}{x^2}+ \cdots \right]\, ,
\label{scaling_function.3}
\end{equation}
in full agreement with the tail in Eq.\ (\ref{fchainint2}). When
$x\to -\infty$, one can use the symmetry relation in Eq.\
\ref{scaling_symmetry} and Eq.\ (\ref{scaling_function.3}) to obtain
the asymptotics. Clearly $F(x)\to 1$ as $x\to -\infty$. Furthermore,
when $x\to 0$, we get
\begin{equation}
F(0)= \frac{1}{\pi} \int_0^{\infty} K_0(y)\, dy=\frac{1}{2}\, .
\label{scaling_function_0}
\end{equation}
This is also consistent with the fact that
$n_m(r) \to 1/2$ as $r\to 0$ (where $1/2$ is the average uniform density
attained in the system at long times in the absence of reset).

\subsection{Dirac Fermions}
\label{dirac1}

In this subsection, we shall study a class of periodically driven
integrable models whose Hamiltonian can be represented by free Dirac
fermions in $d$-dimensions:
\begin{eqnarray}
H(t)= \sum_{\vec k} \psi_{\vec k}^{\dagger} H_{\vec k}(t) \psi_{\vec
k}, \label{fermham}
\end{eqnarray}
where $\vec k$ is the $d$-dimensional momentum, $\psi_{\vec k}=
(c_{\vec k}, c_{-\vec k}^{\dagger})^T$ is the two-component
fermionic field, $c_{\vec k}$ are the annihilation operators for the
fermions, and $H_{\vec k}(t)$ is given by
\begin{eqnarray}
H_{\vec k} = (g(t) - b_{\vec k}) \tau^z + (\Delta_{\vec k}
\tau^+ + \rm{H.c.} ). \label{fermhamden}
\end{eqnarray}
Here $g(t)$ is a periodic function of time characterized by a time
period $T$, and $\Delta_{\vec k}$ and $b_{\vec k}$ can be arbitrary
functions of momenta whose specific forms depend on the particular
physical system which the model (Eq.\ \ref{fermham}) represents. We
note here that Eq.\ (\ref{fermham}) may represent several integrable
models such as the Ising and $XY$ models in $d=1$
[\onlinecite{subirbook1}], the Kitaev model in $d=2$
[\onlinecite{kit1}], triplet and singlet superconductors in $d>1$,
and Dirac fermions in graphene and atop topological insulator
surfaces [\onlinecite{netorev,topoinrev}]. In what follows, we shall
first obtain general results by analyzing fermionic systems given by
Eq.\ (\ref{fermhamden}). The relevance of these results in the
context of specific models will be discussed later in the section.

To study the dynamics of these periodically driven integrable models
with resets, we choose the following protocol. We draw a random
integer, $n$, from a distribution $P_r(n)$ of our choice
characterized by the reset rate $r$ and let the system evolve for
$n$ drive cycles starting from an initial state $|\psi_0\rangle$.
After this, we measure correlation functions of the system. This is
followed by a reset to the initial state $|\psi_0\rangle$. This
process is repeated for several times and correlation functions are
averaged over all measurements. The specific correlation functions
used in our study shall be discussed in details later in this
section.

To analyze the behavior of the correlation function in such
dynamics, we first note that since these models are Gaussian, we
would need to study only the quadratic fermionic correlators. The
states of the system at $t=0$ and for any given $\vec k$  is given
by
\begin{eqnarray}
|\psi_{\vec k}\rangle = \left(\begin{array}{c} u_{\vec k} \\
v_{\vec k} \end{array} \right),\quad |\psi\rangle = \prod_{\vec k >0
} |\psi_{\vec k}\rangle
\end{eqnarray}
where $u_{\vec k}[v_{\vec k}] = (1-[+] (g(0)-b_{\vec k})/E_{\vec
k})^{1/2}/\sqrt{2}$, and $E_{\vec k}= \sqrt{(g(0)-b_{\vec k})^2
+\Delta_{\vec k}^2}$. For simplicity we shall start from an initial
state $(u_{\vec k}, v_{\vec k})=(0,1)$ and drive the system
according to some periodic protocol with time period $T$ for $n$
periods. The quadratic fermionic correlators of the model are given
by
\begin{eqnarray}
C_{\vec k}(nT) &=&  \langle \psi_{\vec k} (nT)|c_{\vec
k}^{\dagger} c_{\vec k}|\psi_{\vec k}(nT)\rangle = |v_{n \vec k}|^2 \label{matrices1}  \\
F_{\vec k} (nT)&=& \langle \psi_{\vec k}(nT)|c_{\vec k} c_{-\vec
k}|\psi_){\vec k}(nT)\rangle= v^*_{n \vec k}(t) u_{n \vec k}
\nonumber
\end{eqnarray}
where $|\psi_{\vec k}(nT)\rangle = (u_{n \vec k},v_{n \vec k})^T$ is
the wavefucntion of the system at momentum $\vec k$ and after $n$
drive cycles.

Next we note that for any periodic drive, the time periodicity of
the Hamiltonian ensures that the unitary evolution operator at the
end a drive cycle can be written as $U(0, T)= \exp[-i H_F T]$ where
$H_F$ is the Floquet Hamiltonian [\onlinecite{floquetrev1}].
Consequently, the correlation functions of such driven systems, at
the end of $n$ drive periods, can also be expressed in terms of
eigenfunctions and eigenvectors of $H_F$
[\onlinecite{floquetrev1,asen1}]. For the class of integrable models
that we treat here, one can show [\onlinecite{asen1}]
\begin{eqnarray}
U_{\vec k} &=& \left(\begin{array}{cc}
\cos(\theta_{\vec k}) e^{i \alpha_{\vec k}} & \sin(\theta_{\vec k})e^{i \gamma_{\vec k}} \\
-\sin(\theta_{\vec k})e^{-i \gamma_{\vec k}}& \cos(\theta_{\vec k}) e^{-i \alpha_{\vec k}} \\
\end{array} \right)
= e^{-i H_{\vec{k}F} T}  \label{ueq1a}\nonumber\\
\end{eqnarray}
where the parameters $\theta$, $\alpha$ and $\gamma$ can be found in
terms of initial [$(u_0,v_0)=(0,1)$] and final [$(u_{\vec k}, v_{
\vec k})$] wavefunctions after one drive cycle as
\begin{figure}
\includegraphics[width=\linewidth]{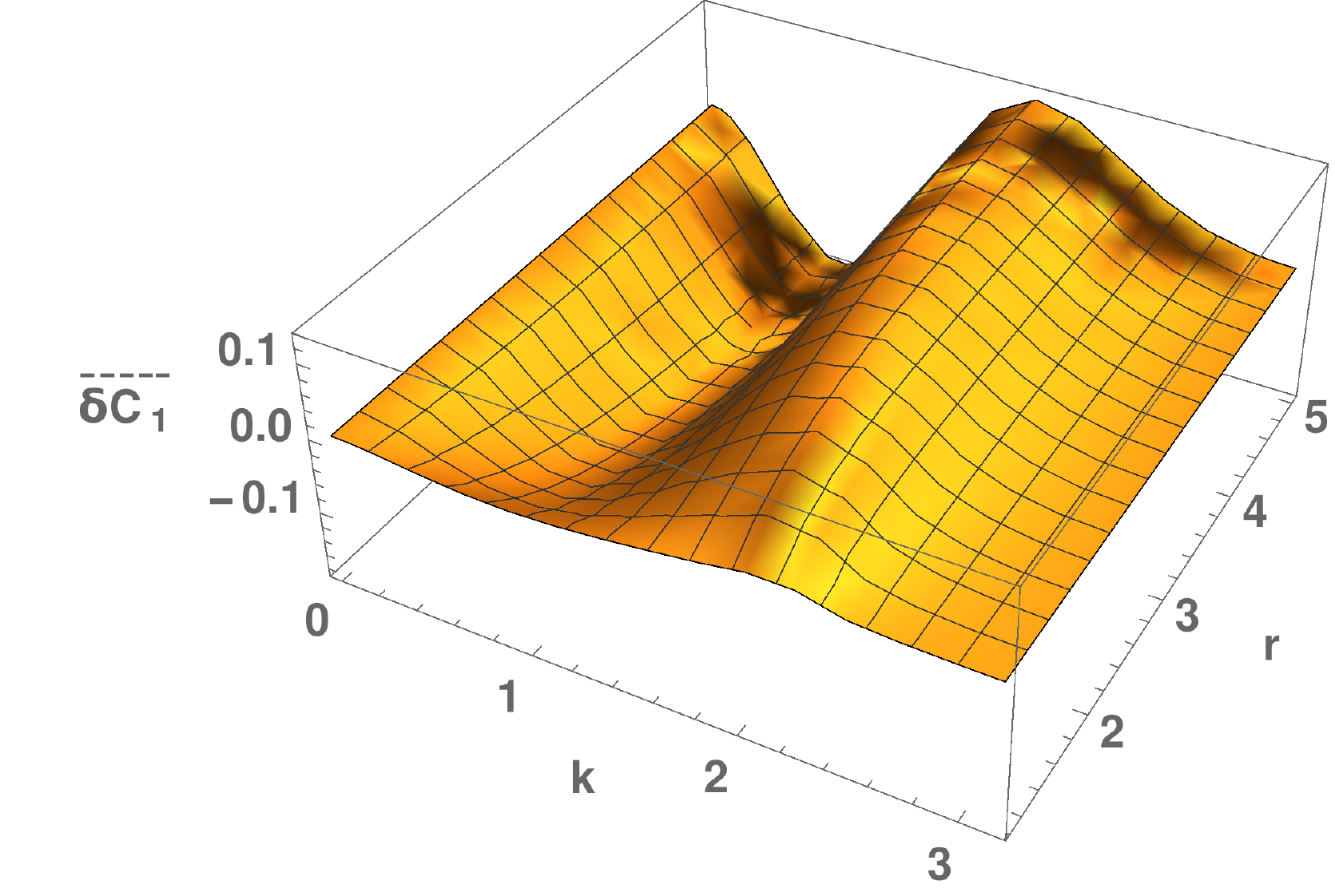}
%\rotatebox{0}{
%\includegraphics*[bb=0 0 246 238, width=\linewidth]{Fignew.jpg}}
\caption{Plot of $\overline{\delta C_1}$ as a function of $k$ and
$r$ for the pulse protocol with $g_0=2$, $g_1=0.5$ and $T=1$. All
energies are scaled in units of $J$.} \label{fig3}
\end{figure}
\begin{eqnarray}
\sin(\theta_{\vec k})&=& |u_{\vec k}|,\quad \alpha_{\vec k}
(\gamma_{\vec k}) = {\rm Arg}[u_{\vec k}(v_{\vec k})] \label{ueq1b}
\end{eqnarray}
For more general choice of the initial wavefunctions, the
expressions for $\theta$, $\alpha$ and $\gamma$ can be found in
Ref.\ \onlinecite{asen1}. Furthermore, since $U_{\vec k}$ is a SU(2)
matrix, one gets [\onlinecite{asen1}]
\begin{eqnarray}
\label{ueq3} U_{\vec k} &=& e^{-i (\vec \sigma \cdot \vec n_{\vec
k}) \phi_{\vec k}}, \quad n_{\vec k}= \frac{\vec \epsilon_{\vec
k}}{|\vec \epsilon_{F \vec k}|} \quad, \phi_{\vec k} = T|\vec
\epsilon_{F \vec k}| \label{ueq2a}
\end{eqnarray}
where
\begin{eqnarray}
n_{\vec{k}1} &=& -\sin(\theta_{\vec k}) \sin(\gamma_{\vec
k})\sin(\phi_{\vec k})/D_{\vec k}
\nonumber\\
n_{\vec{k}2} &=& -\sin(\theta_{\vec k}) \cos(\gamma_{\vec
k})\sin(\phi_{\vec k})/D_{\vec k}
\nonumber\\
n_{\vec{k}3} &=& -\cos(\theta_{\vec k}) \sin(\alpha_{\vec
k})\sin(\phi_{\vec k})/D_{\vec k}
\nonumber\\
D_{\vec k}&=& \sqrt{1-\cos^2(\theta_{\vec k}) \cos^2(\alpha_{\vec
k})} \nonumber\\
|\vec \epsilon_{F \vec k}| &=&   \arccos[\cos(\theta_{\vec k})
\cos(\alpha_{\vec k})]/T.  \label{ueq3a}
\end{eqnarray}
Here ${\rm Sgn}$ denotes the signum function. Note that at the edge
of the Brillouin zone, where the off-diagonal component of $H_k $
disappears, $U_k$ becomes a diagonal matrix, which in turns makes
$\sin(\theta_{\vec k })=0$. This leads us to the result
$n_{\vec{k}1}= n_{\vec{k}2}=0$ and $n_{\vec{k}3}=\pm 1$ for such
momentum values.

Using the fact $|\psi_{\vec k}(nT)\rangle = U^n_{\vec k} |\psi_{\vec
k}(0)$ and Eqs.\ (\ref{matrices1}..\ref{ueq3a}), after some algebra,
we can write
\begin{eqnarray}
\delta C_{\vec k}(n) &=&    f_1(\vec{k})\cos(2n\phi_{\vec k}) \label{generalresults} \\
 \delta F_{\vec k}(n) &=&  (f_2(\vec{k})\cos(2n\phi_{\vec k}) +f_3(\vec{k})\sin(2n\phi_{\vec k})) \nonumber
\end{eqnarray}
where  $\delta C_{\vec k}(n)= \langle c_{\vec k}^{\dagger} c_{\vec
k} \rangle_n - \langle c_{\vec k}^{\dagger} c_{\vec k}
\rangle_{\infty}$ and similarly for $\delta F_{\vec k}(n)$. In Eq.\
(\ref{generalresults}), the quantities $f_1(\vec k)$, $f_2(\vec k)$,
and $f_3(\vec k)$ are given in terms of elements of the Floquet
Hamiltonian $H_F$ as
\begin{eqnarray}
f_1(\vec{k}) &=& -(1-n^2_{\vec{k}3}), \quad  f_2(\vec{k}) = -i
\hat{n}_{\vec{k}3}
f_3(\vec{k}) \nonumber\\
f_3(\vec{k}) &=& i({n}_{\vec{k}1}+i{n}_{\vec{k}2}). \label{ffns}
\end{eqnarray}
Note that $\delta C$ and $\delta F$ vanishes by construction for $n
\to \infty$. The contribution to these terms comes from the
off-diagonal terms of the density matrix in the Floquet basis as is
evident from the presence of $\cos(2n\phi_{\vec k}) $ and
$\sin(2n\phi_{\vec k})$ factors in their expression
[\onlinecite{asen1}]. For $n \to \infty$, the system is described by
a generalized Gibbs ensemble (GGE) which is characterized by the
values of the correlations $C_{\infty}(\vec k)= 1-n_{\vec k 3}^2$
and $F_{\infty}(\vec k)= -n_{\vec k 3}(n_{\vec k 1}+ i n_{\vec k
2})$. Any deviation of the value of the fermionic correlators from
$C_{\infty}$ or $F_{\infty}$ in the steady state therefore
constitutes a different GGE representing that state. For this to
happen, one clearly needs non-zero values of $\delta C$ or $\delta
F$ for such steady states.

\begin{figure}
\includegraphics[width=\linewidth]{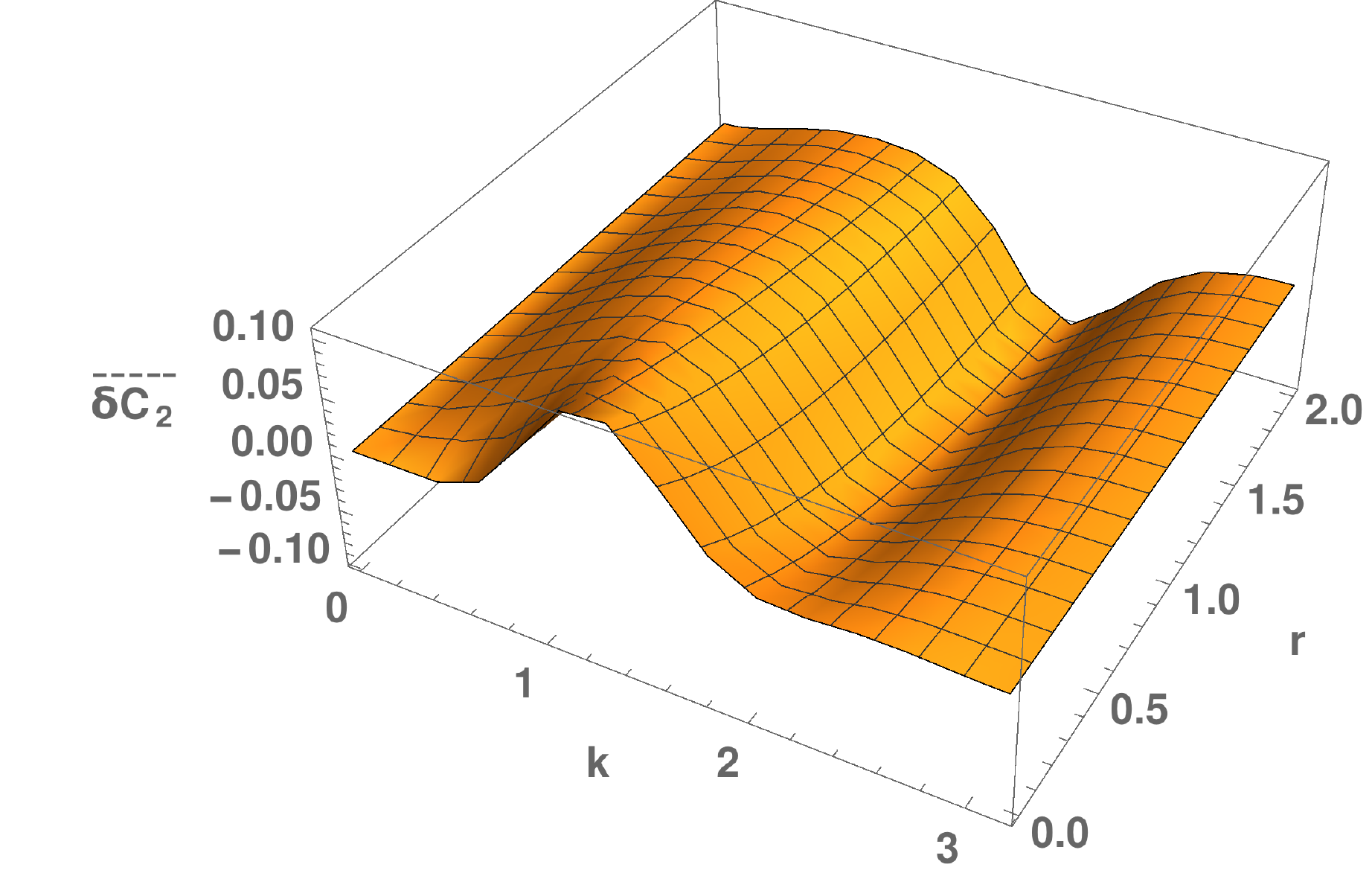}
%\rotatebox{0}{
%\includegraphics*[bb=0 0 246 238, width=\linewidth]{Fignew.jpg}}
\caption{Plot of $\overline {\delta C}_2$ as a function of $k$ and
$r$. All parameters are same as in Fig.\ \ref{fig3}.} \label{fig4}
\end{figure}

Next, we look into the evolution of such a periodically driven model
in the presence of resets characterized by $P_r(n)$ as discussed
earlier. We note here that the resets do not mix the different $\vec
k$ modes and hence retain the underlying integrability of the model.
Thus we do not expect the steady-state density matrix to be given by
a Gibbs ensemble with a characteristics temperature. The average
value of $\delta C_{\vec k}$ and $\delta F_{\vec k}$, under such a
reset protocol, is given by
\begin{eqnarray}
\overline{\delta C_{\vec k}}(r) = \sum_{n=1}^{\infty} \delta C_{\vec
k}(n) P_r(n) \nonumber\\
\overline  {\delta F_{\vec k}}(r) = \sum_{n=1}^{\infty} \delta
F_{\vec k}(n) P_r(n)
\end{eqnarray}
Note that a finite value of ${\overline  \delta C_{\vec k}}(r)$ or
${\overline \delta F_{\vec k}}(r)$ for any $k$ signifies a GGE
characterized by $r$ which is different from the one at $r=0$. In
what follows we shall focus on $\delta C$.

Below, we choose two different probability distributions for
$P_r(n)$. The first one is given by $P_{r}(n) = [{\rm Li}_r(1)]^{-1}
1/n^r$ for $r>1$ and $ r \in Z$, where ${\rm Li}_r[1] =\sum_n
(1/n^r)$ is the PolyLog function. In the first case one obtains
\begin{eqnarray}
\overline {\delta C_{1\vec k}}(r) &=& f_1(\vec k) {\rm Re}[ {\rm
Li}_r( e^{ 2 i \epsilon_{F \vec k} T})]/{\rm Li}_r[1] \label{corr1}
\end{eqnarray}
where $\epsilon_{F \vec k}$ is the Floquet energy spectrum (Eq.\
\ref{ueq3a}).

The second one is the more well-known Poisson distribution for which
$ P_r(n)= r^n \exp[-r]/n!$. For this, we find
\begin{eqnarray}
\overline {\delta C_{2\vec k}}(r) &=& f_1(\vec k) [ e^{-r(1-\cos(2
\epsilon_{F \vec k} T))} \nonumber\\
&& \times \cos[\sin(2 \epsilon_{F \vec k}T)] - e^{-r}] \label{corr2}
\end{eqnarray}

To check if $\overline{\delta C_1}$ (Eq.\ \ref{corr1}) and
$\overline{\delta C_2}$ (Eq.\ \ref{corr2}) are finite functions of
$r$ and $k$, we consider one-dimensional Ising model in a transverse
field. The Hamiltonian of this model can be mapped to Eq.\
(\ref{fermham}) with $b_k/J= \cos(k)$ and $\Delta_k/J = \sin(k)$,
where $J$ is the interaction strength between neighboring Ising
spins and $g=h/J$ is the strength of the transverse field. We plot
$\overline{\delta C_1}$ and $\overline{\delta C_2}$ as a function of
$k$ and $r$. We choose a periodic delta function pulse protocol
$g(t)= g_0 + g_1 \sum_n \delta (t-nT)$ with $g_0=2$, $g_1=0.5$ and
$T=1$. For this protocol, $\epsilon_{F \vec k}= \arccos[ \cos
(E_{\vec k}T) +(1-n_{3 \vec k }^2) \sin(E_{\vec k}T)]/T$
[\onlinecite{asen1}]. Substituting this in Eqs.\ (\ref{corr1}), and
(\ref{corr2}), one may obtain $\delta C_{1(2)\vec k}$ as a function
of $g_0$, $g_1$, $T$, $r$ and $\vec k$. The results, shown in Fig.\
\ref{fig3} for $1/n^r$ distribution and Fig.\ \ref{fig4} for the
Poisson distribution, clearly indicates that both $\overline{\delta
C_1}$ and $\overline{\delta C_2}$ are finite and functions of $r$
for all $k \ne 0, \pi$ where these correlations are identically zero
(since $f_{1,2}=0$ and $f_3=1$ for these momenta). This clearly
shows that stochastic resets lead to distinct family of GGEs
characterized by a reset rate $r$ for periodically driven integrable
quantum systems.

\section{Bose-Hubbard model in a tilted optical lattice}
\label{tbh}

In this section, we consider the stochastic dynamics of a Bose
Hubbard model in the presence of a tilt, or an effective electric
field. To see how such an electric field can be realized, first let
us consider a typical Bose Hubbard model in a deep one-dimensional
(1D) optical lattice (with lattice spacing $a$) so that the bosons
are localized with $n=n_0$ bosons occupying each lattice site. The
boson system is described by the well-known Bose-Hubbard Hamiltonian
given by
\begin{eqnarray}
H &=& -J \sum_{\langle r r' \rangle} (b_r^{\dagger} b_{r'} + {\rm
h.c.} )  \nonumber\\
&& + \sum_{r} (-\mu n_{r} + U n_r(n_r-1)/2) \label{bhm1}
\end{eqnarray}
where  $J$ is the nearest neighbor hopping amplitude of the bosons,
$U$ is the on-site interaction potential, and $\mu$ is the chemical
potential. Here $b_r$ denotes the boson annihilation operator at
site $r$, $n_r = b_r^{\dagger} b_r$ is the boson number operator,
and $\langle .. \rangle$ denotes sum over nearest neighboring sites
of the lattice. We choose $J/U \ll 1$ and $\mu= \mu_0$ so that the
ground state of $H$ represent a Mott localized state of bosons with
$n_0$ bosons per site.

To generate a tilt for the bosons, the most experimentally
convenient way is to apply a Zeeman magnetic field with varies
linearly in space: $B(r)=B_0 (r/a)$. The Zeeman term for this bosons
can be written as $H_z= -\sum_r g \mu_B B_0 (r/a) n_{r} = \sum_r
{\mathcal E} r n_r$, where ${\mathcal E}= g \mu_B B_0/a$ is the
effective electric field seen by the bosons, $B_0$ is the field
amplitude, and $\mu_B$ is the Bohr magneton. The Hamiltonian of the
system in the presence of the tilt is given by
\begin{eqnarray}
H &=& -J \sum_{\langle r r' \rangle} (b_r^{\dagger} b_{r'} + {\rm
h.c.} )  \nonumber\\
&&+ \sum_{r} (-(\mu +{\mathcal E} r) n_{r} + U n_r(n_r-1)/2)
\label{bhm2}
\end{eqnarray}
The equilibrium and non-equilibrium properties of this Hamiltonian
has been studied in several situations
[\onlinecite{subir1,dynamicsdipole}]. To understand the property of
such a system, it is first useful to note that a system of
non-interacting bosons ($U=0$) in the presence of a tilted optical
lattice constitutes a Wannier-Stark problem with exponentially
localized wavefunctions. Thus, contrary to the classical
expectation, bosons do not move to the last site to minimize their
energy. Such a movement which constitutes an electric breakdown
involves tunneling of the bosons to higher single particle bands.
The time required for such breakdown for ultracold boson systems
turns out to be larger than the system lifetime. Thus the parent
boson state is preserved within experimental timescales. This
feature is preserved, albeit with some difference, in the Mott
regime where $U$ is large. The strategy for a theory of such a
system thus involves identifying the low energy subspace around the
parent Mott state which, in the presence of the electric field, is a
metastable state with a very long lifetime [\onlinecite{subir1}].

It turns out that the low-energy theory of such a state can be
formulated in terms of dipoles [\onlinecite{subir1}]. The creation
of a dipole involves hopping of a boson from a site of the 1D
lattice to its next neighbor. This costs an energy $U- {\mathcal
E}$. Thus in the parameter regime where $U-{\mathcal E}, J \ll U,
{\mathcal E}$, the low-energy effective Hamiltonian of the bosons in
a tilted lattice can be described in terms of dipole operators
$d^{\dagger}_{\ell} = b_j^{\dagger} b_i/\sqrt{n_0(n_0+1)}$, where
$\ell$ denotes the link between sites $i$ and $j$, as
\begin{eqnarray}
H_d &=& -w \sum_{\ell} (d_{\ell} + d_{\ell}^{\dagger}) +
(U-{\mathcal E}) \sum_{\ell} n_{\ell}.  \label{diham}
\end{eqnarray}
Here $n_{\ell} = d_{\ell}^{\dagger} d_{\ell}$ is the dipole number
operator and $w= J \sqrt{n_0(n_0+1)}$. The dipole model, so
constructed, has two constraints. First, there can not be more than
one dipole in a given site ($n_{\ell} \le 1$) and second, there can
not be two dipoles on adjacent links ($n_{\ell} n_{\ell + 1} =0$)
[\onlinecite{subir1}]. These constrains arise as the states which do
not obey them can be shown not to be a part of the low-energy
subspace with respect to the parent Mott state. It has been shown
that the dipole model leads to two distinct ground states. The first
is the dipole vacuum which occurs at $U > {\mathcal E}$; in the
boson language this corresponds to the parent Mott state with $n_0$
bosons per sites. The second is the maximal dipole ground state
occurring at ${\mathcal E} \ge {\mathcal E}_c = U + 1.31 w$ which
corresponds to a $Z_2$ symmetry broken state with a dipole on odd or
even links (but not both due to the second constraint mentioned
above). In terms of the original boson model this state corresponds
to $n_0+1$ and $n_0-1$ bosons on every alternate site. These two
states are separated by a quantum phase transition belonging to the
Ising universality class at ${\mathcal E}= {\mathcal E_c}$. The
quantum dynamics of the model by sudden, ramp and periodic time
variation of the electric field has been studied in Ref.\
\onlinecite{dynamicsdipole}.

\begin{figure}
\includegraphics[width=0.49\linewidth]{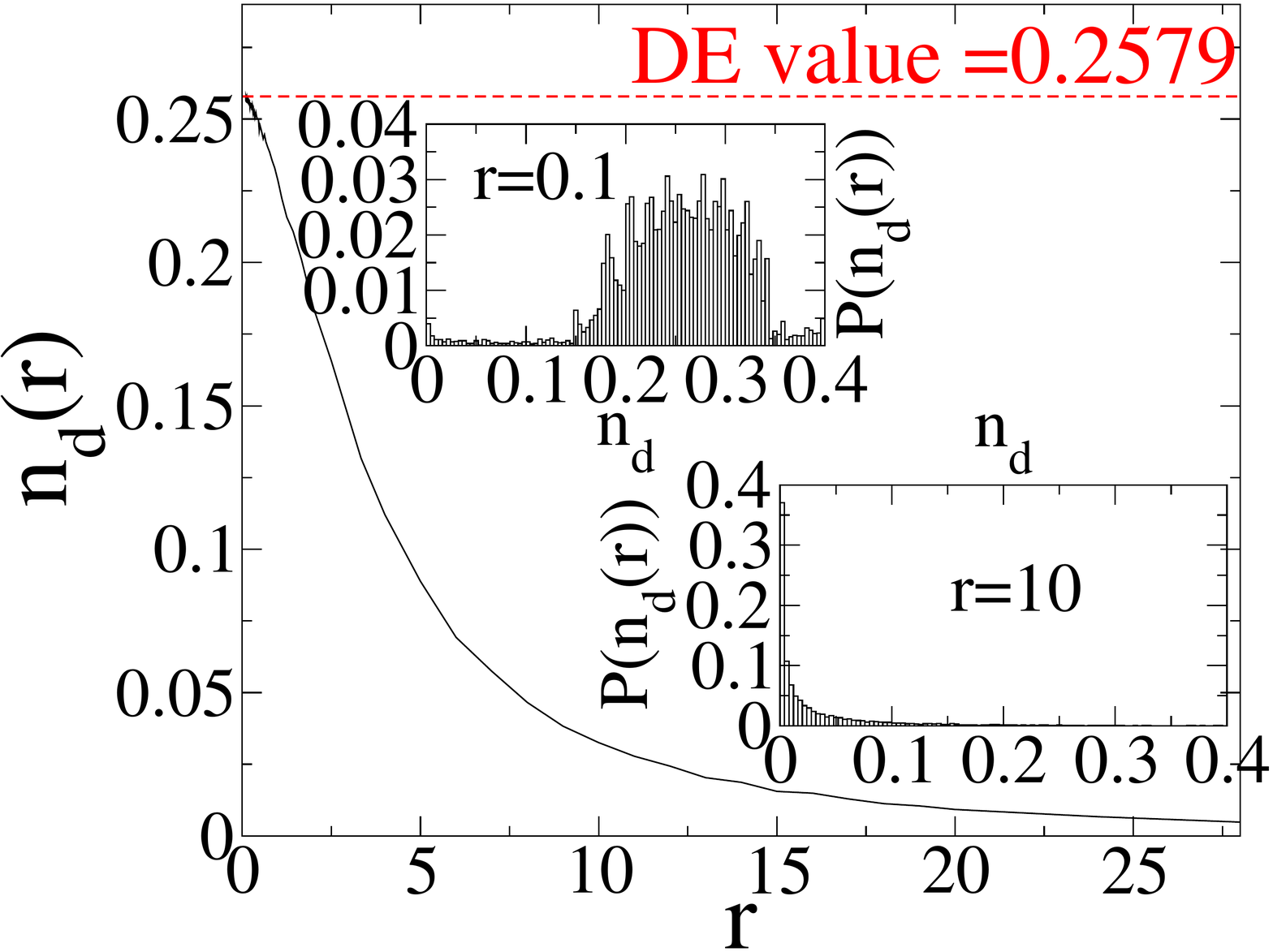}
\includegraphics[width=0.49\linewidth]{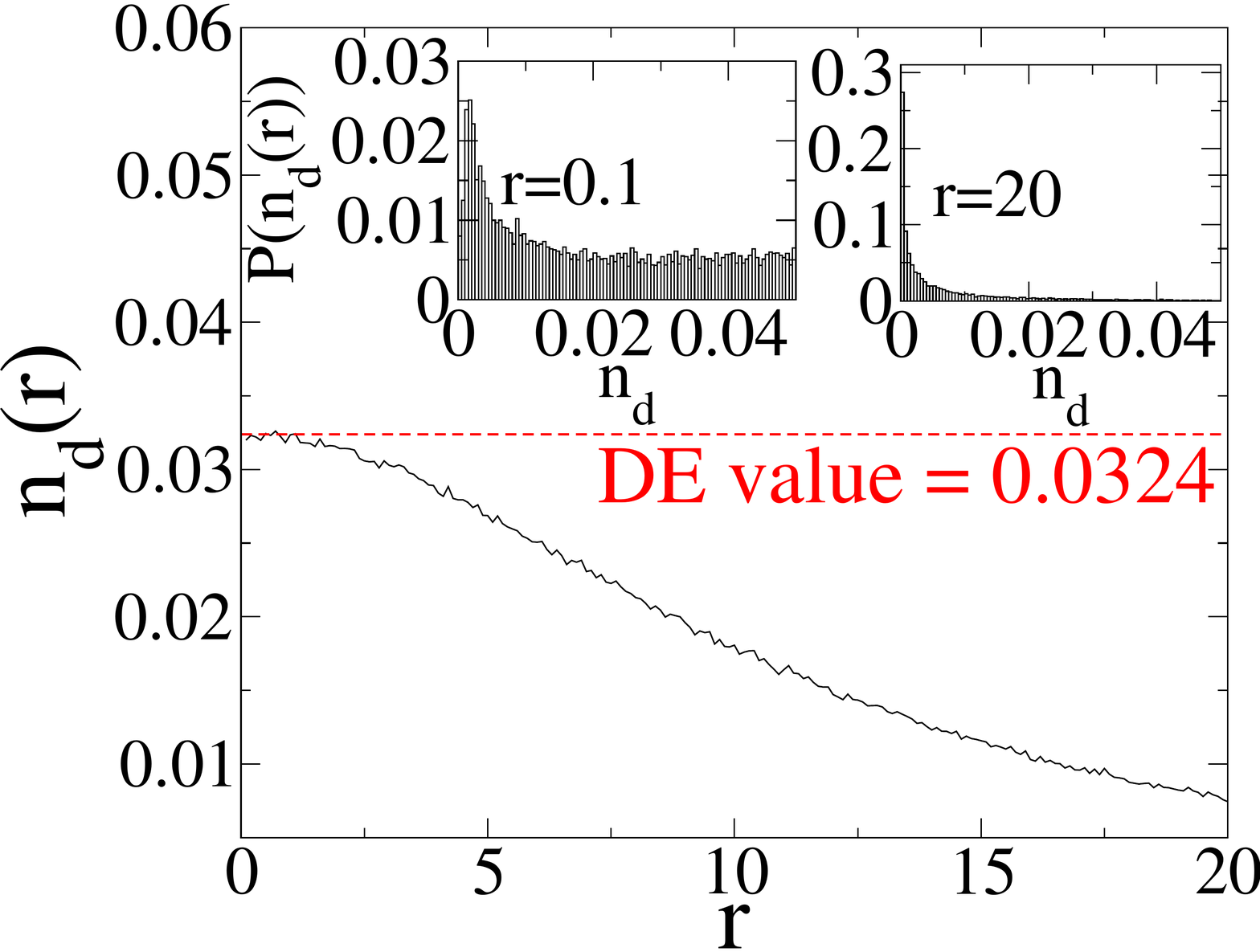}
%\rotatebox{0}{
%\includegraphics*[bb=0 0 246 238, width=\linewidth]{Fignew.jpg}}
\caption{Plot of the dipole density $n_d(r)$ as a function of $r$
for (a) $U=E$ and (b) $(U-E)/w=-10$. The inset shows the probability
distribution of $n_d$ at fixed $r$. The plots correspond to
$N_0=10000$ and $N=12$. } \label{fig5}
\end{figure}

Below, we compute the steady state expectation value of the dipole
density operator $n_d = \sum_{\ell} n_{\ell} /N$, where $N$ is the
number of lattice sites in the chain and we have set the lattice
spacing to unity. We start from a dipole vacuum state which is the
ground state of the system for $U > {\mathcal E}$ and $w=0$ and
study its evolution under the dipole Hamiltonian $H_d$ with
$(U-{\mathcal E}_f)/w=0$ and $(U-{\mathcal E}_f)/w=-10$ using exact
diagonalization (ED). The former parameter corresponds to the system
being near the critical point while the latter corresponds a maximal
dipole ground state. The initial state of the system is denoted by
$|\psi_0\rangle$ for which $\langle n_d \rangle=0$. The state of the
system can be expressed at any instant $t>0$ as
\begin{eqnarray}
|\psi(t)\rangle &=& \sum_{\alpha} c_{\alpha} e^{-i \epsilon_{\alpha}
t} |m\rangle
\nonumber\\
H[{\mathcal E}_f] |\alpha \rangle &=& \epsilon _{\alpha}
|\alpha\rangle \quad c_m= \langle m|\psi_0\rangle \label{evol1}
\end{eqnarray}
where $\epsilon_{\alpha}$ and $|\alpha\rangle$ can be obtained
numerical diagonalization of $H[{\mathcal E}_f]$. We note that since
$H_d$ can be represented by a  real symmetric matrix, $c_{\alpha}$
can be chosen to be real. We shall use this choice for the rest of
the section. Using Eq.\ (\ref{evol1}), one obtains
\begin{eqnarray}
n_d(t) &=& \frac{1}{N} \sum_{\alpha \beta} c_{\alpha}  c_{\beta}
e^{-i \omega_{\beta \alpha} t} \langle \alpha|\sum_{\ell} n_{\ell}
|\beta \rangle, \label{ndtime}
\end{eqnarray}
where $\omega_{\beta \alpha}= (\epsilon_{\beta}-\epsilon_{\alpha})$.
Note that $n_d(t=0)=0$ since the initial state corresponds to a
dipole vacuum. In contrast, the steady state (diagonal ensemble)
value is finite and is found numerically to be $n_d(t \to
\infty)=n_d^{\rm de} =0.2575$ for $U={\mathcal E}$. The
corresponding value for $(U-{\mathcal E})/w =-10$ is $n_d^{\rm
de}=0.0324$. The reason for a smaller value of $n_d^{\rm de}$ in the
ordered phase can be understood as follows
[\onlinecite{dynamicsdipole}]. First, we note that deep in the
ordered phase $|\psi_0\rangle$ has substantial overlap with only a
few of the eigenstates of the Hamiltonian; these eigenstates
corresponds to $\langle n_d \rangle \simeq 0$. Thus $n_d^{\rm de}
\simeq \sum_{\alpha} c_{\alpha}^2 \langle \alpha| \hat n_d
|\alpha\rangle$ remains small. In contrast, near the critical point,
$|\psi_0\rangle$ has finite overlap with several eigenstates of the
near-critical Hamiltonian leading to a larger value of $n_d^{\rm
de}$.

\begin{figure}
\includegraphics[width=\linewidth]{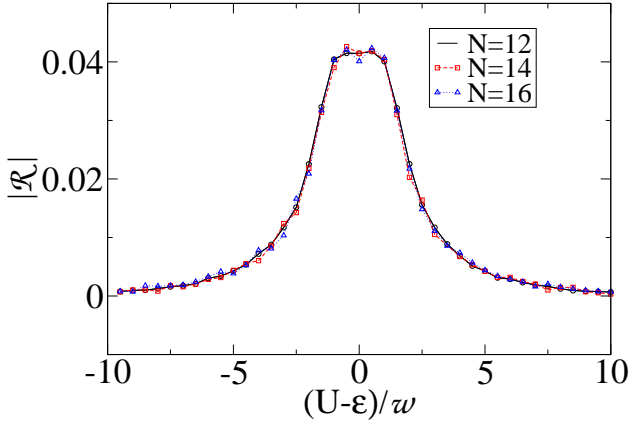}
%\rotatebox{0}{
%\includegraphics*[bb=0 0 246 238, width=\linewidth]{Fignew.jpg}}
\caption{Plot of ${\mathcal R}$ as a function of $(U-{\mathcal
E})/w$ for $N=12, \, 14$, and $16$. } \label{fig6}
\end{figure}

Next, we modify the unitary evolution following the quench with
stochastic reset of the system at a time $\tau$ which is a random
number with $P_r(\tau)= r \exp[-r \tau]$. Our numerical strategy for
finding the effect of the reset is the following. First, we let the
system evolve up to a time $\tau$ which is obtained using a random
number generator and measure $n_d(\tau_j|t_0) \equiv n_d(\tau_j)$ at
a fixed large $t_0$. We repeat this process for $N_0=10000$ and
obtained average $n_d(r) = \sum_{j=1}^{N_0} P_r(\tau_j) n_d(\tau_j)$
from the data. Finally we repeat the procedure for several $r$. Note
that since we can ignore the probability of having zero reset by
formally choosing a large measurement time, the numerical procedure
is expected to produce identical result to those obtained via Eqs.\
(\ref{denmat2}) and (\ref{opexp}). We note that we have checked that
for finite $r$, the system retains finite off-diagonal matrix
elements of the density matrix for large $N_0$: $\langle m|
\rho|n\rangle \ne 0$ for $m \ne n$ in the steady state. Thus the
resultant deviation of $n_d$ or $S$ from their $r=0$ steady state
values can not be interpreted as due to a steady state Gibb's
distribution with $r$ dependent temperature.

The results obtained from such a procedure are shown in Fig.\
\ref{fig5}(a) and \ref{fig5}(b) for $U={\mathcal E}$ and
$(U-{\mathcal E})/w=-10$ respectively. We find that in both cases
the steady state values of the dipole density with a fixed reset
rate, $n_d(r)$, interpolates between $n_d(t=0)=0$ (for $r \to
\infty$) and the diagonal ensemble value $n_d^{de}$ (for $r \to 0$).
The $r$ dependence of $n_d(r)$ indicates that the steady state
density matrix retains off-diagonal matrix elements and thus does
not correspond to a diagonal density matrix for any finite $r$. We
also note that the initial decrease of $n_d(r)$ from the diagonal
ensemble value ($r=0$) to the Zeno (initial) value ($r \to \infty$)
is faster for $U={\mathcal E}$. This feature can be qualitatively
understood as follows. We note that the slope of $n_d(r)$ near $r=0$
can be expressed using Eq.\ (\ref{opexp2}) as
\begin{eqnarray}
{\mathcal R} &=& \frac{d n_d(r)}{dr}= \sum_{\alpha > \beta}
c_{\alpha} c_{\beta} \langle \alpha| n_d |\beta\rangle \frac{2
\omega_{\beta \alpha}^2 r}{(r^2+\omega_{\beta \alpha}^2)^2}.
\label{slopeeq}
\end{eqnarray}
where we have used the fact $\langle \alpha|n_d|\beta\rangle =
\langle \beta|n_d|\alpha \rangle$. Note that near the critical point
larger number of states have a finite overlap with $|\psi_0\rangle$
rendering a larger number $c_{\alpha}$s finite; this leads to
enhancement of ${\mathcal R}$ for small but finite $r$. For $r=0$
and $r \to \infty$, the slope vanishes. A plot of $|{\mathcal
R}(r=1)|$ for as a function of $(U-{\mathcal E})/w$ for a
representative $r=1.2$ is shown in Fig.\ \ref{fig6} confirming this
expectation. Thus we find that $|{\mathcal R}|$ for a typical finite
$r$ is sensitive to the presence of a critical point in the system
and is expected to peak around it; however, its peak need not be at
the precise location of the critical point.

\begin{figure}
\includegraphics[width=0.49\linewidth]{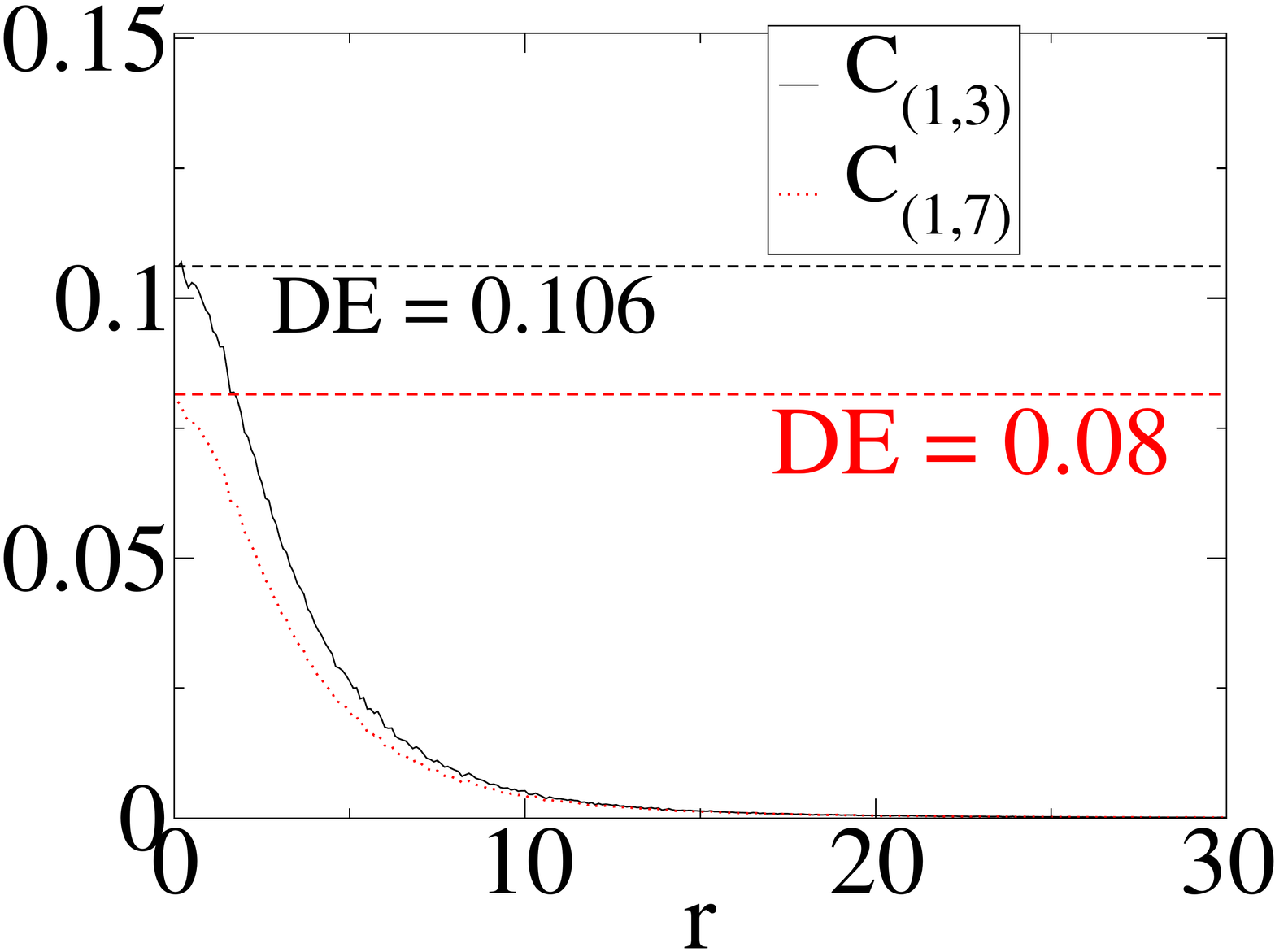}
\includegraphics[width=0.49\linewidth]{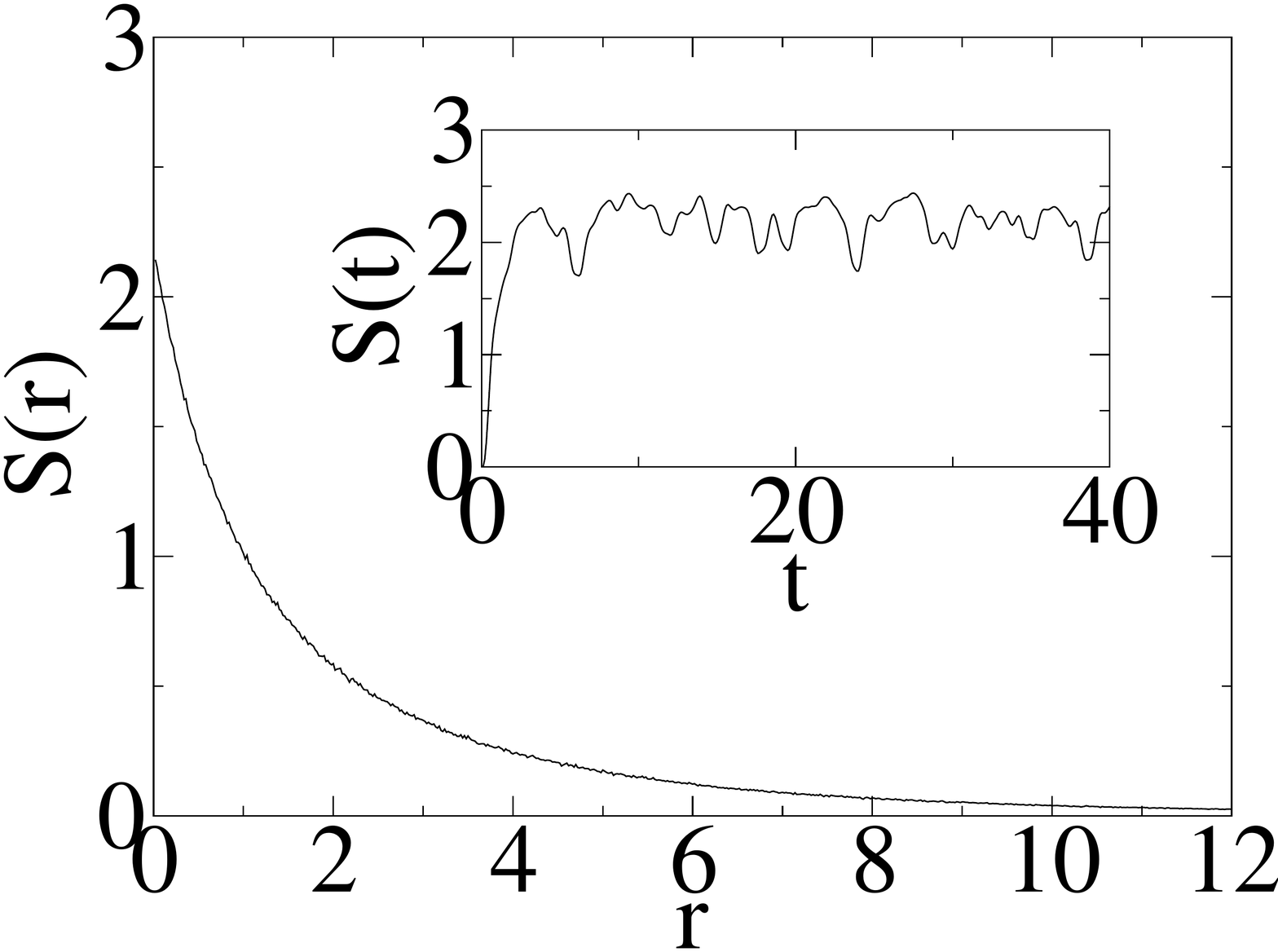}
%\rotatebox{0}{
%\includegraphics*[bb=0 0 246 238, width=\linewidth]{Fignew.jpg}}
\caption{Left Panel: Plot of the dipole density-density correlation
function $C_{\ell \ell'}(r)$ for two representative values of
$|\ell-\ell'|$ as a function of $r$. Right Panel: Plot of the half
chain entanglement $S(r) $ as a function of $r$. The inset shows
$S(t)$ for a particular unitary time evolution between two resets
after the quench. The diagonal ensemble state value of $S=2.18$ as
can be seen from the inset. Both the plots correspond to
$N_0=10000$, $U-{\mathcal E}/w=0$, and $N=12$. } \label{fig7}
\end{figure}

Next we compute the dipole correlation function $C_{\ell,\ell'}(r) =
\langle \langle n_d(\ell) n_d(\ell') \rangle \rangle$ where the
expectations correspond to that with respect to $|\psi(\tau)\rangle$
and average over $P_r(\tau)$. The result is shown in Fig.\
\ref{fig7}(a) for $\ell=1$ and $\ell'= 3$ and $\ell'=7$. The
diagonal ensemble or the steady state value of this correlation
function is shown via dotted horizontal lines in Fig.\
\ref{fig7}(a). Once again we find that the value of $C_{\ell
\ell'}(r)$ interpolates between its diagonal ensemble value for $r
\to 0$ and the initial value for $ r \to \infty$. In Fig.\
\ref{fig7}(b), we plot the half-chain entanglement
\begin{eqnarray}
S(r) &=& \int_0^{\infty}  dt r e^{-rt } S(t), \,\, S(t)=-{\rm Tr}
[\rho(t) \ln \rho(t)] \nonumber\\ \label{entangeq}
\end{eqnarray}
where the reduced density matrix $\rho$ for $N/2$ sites in the chain
is computed numerically using ED. We note that $S(r)$ also
interpolates between the diagonal ensemble $S(0)= S_d \simeq 2.52$
and the quantum Zeno (initial) $S(\infty) = S_{\rm initial} =0$
values. This confirms our expectation that the reset averaged steady
state density matrix retains off-diagonal matrix elements.

\begin{figure}
\includegraphics[width=\linewidth]{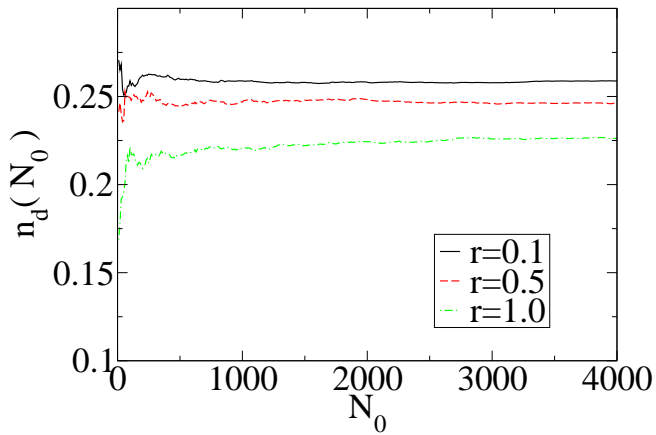}
%\rotatebox{0}{
%\includegraphics*[bb=0 0 246 238, width=\linewidth]{Fignew.jpg}}
\caption{Plot of $n_d(N_0)$ as a function of the number of
measurements $N_0$ showing $N_0$ independence of $n_d$ for large
$N_0$. The plots correspond to $U-{\mathcal E}/w=0$, and $N=12$. See
text for details. } \label{fig10}
\end{figure}

Before ending this section, we show that the reset indeed leads to a
steady state as discussed earlier. To this end, we plot the average
value of the dipole density $n_d (N_0)$ over $N_0$ measurements as a
function of $N_0$ in Fig.\ \ref{fig10} for several representative
values of $r$. We find that $n_d(N_0)$ indeed approaches a constant
value as $N_0$ increases for any $r$ showing that the system reaches
a steady state for large $N_0$. We have checked that a similar
steady-state behavior is displayed by the correlation functions
$C_{\ell \ell'}$.

\section{Discussion}
\label{diss}

In this work, we have studied the unitary dynamics of quantum
integrable and non-integrable systems interrupted by stochastic
resets characterized by a rate $r$. For non-integrable models, such
dynamics leads to a non-thermal steady state density matrix. Our
analysis shows that such a dynamics leads to novel steady states for
non-integrable and GGEs /statianory states for integrable quantum
systems.

For non-integrable systems, we find that the density matrix,
averaged over reset probability distribution, retains off-diagonal
elements in the eigenbasis of the Hamiltonian controlling its
unitary evolution. Thus such dynamics lead to non-diagonal steady
state density matrices. This also indicates that the expectation
value of a generic observable, averaged over the reset distribution,
retain contribution from such off-diagonal terms. Thus the steady
state value of the observable is no longer given by its average over
a diagonal steady state density matrix as is customary for evolution
without reset. We verify these results by explicit numerical
calculation on Bose-Hubbard model on a tilted optical lattice which
has been realized experimentally [\onlinecite{greiner1}]. We compute
the dipole density, dipole density-density correlation and
half-chain entanglement entropy of the model as a function of $r$.
For all computed quantities, we find our results to agree with the
diagonal ensemble result at $r=0$ and the quantum Zeno result for $r
\to \infty$. In between, these quantities are smooth functions of
$r$ indicating the finite contribution of the off-diagonal elements
of the density matrix.

For integrable model, we have studied two separate examples. The
first involves a spinless fermionic chain described by a hopping
Hamiltonian with all the fermions occupying the left half of the
chain. We study the behavior of the fermionic density $n_m(r)$ as a
function of the reset rate and derive a scaling function which
describes its stationary state behavior. Our results predicts a
decay of the fermionic density as a function of the site index and
reset rate as $n_m(r) \sim \exp[-m r]/\sqrt{2 \pi m r}$ for $m r \gg
1$. The second example involves periodically driven Dirac fermions
in the presence of resets. Here we have adapted a protocol where the
system is allowed to evolve under a periodic drive for $n$ cycles
where $n$ is a random integer chosen from a pre-determined
distribution $P_r(n)$. After this evolution, the correlation
functions of the system is computed and its state is reset to its
initial value. This process is repeated and the average correlation
is obtained summing over all measured values weighted by $P_r(n)$.
This yields the steady state correlation function values for a
finite $r$. We demonstrate that these values depend on $r$
indicating that the system is described by GGEs characterized by
$r$.

Experimental verification of our results involves implementing the
reset protocol. We note that such resets can be implemented by
suitable projections of the quantum state to its initial value in
ultracold atom systems by suitable laser pulses. For examples, such
experiments have already been carried out leading to observation of
the  quantum Zeno effect. Such observations constituted experimental
implementation of resets with $ r \gg 1$ [\onlinecite{mukund1}]. We
also note that finite chain of bosons in a tilted lattice have been
experimentally realized [\onlinecite{greiner1}]. In such
experiments, the dipole density computed in our work can be directly
measured via parity of occupation measurement [\onlinecite{greiner2,
greiner1}]. For experimental purpose, we would like to suggest
measurement of bosonic dipole density as a function of $r$. For
this, one would need to reset the system to the dipole vacuum state
with a finite rate. This can in principle be done by changing the
value of the electric field to a small value and letting the system
equilibrate to the ground state of the resultant Hamiltonian
[\onlinecite{greiner1}]. We predict that the reset averaged value of
the dipole density would be a smooth decaying function of $r$ and
would be qualitatively similar to that shown in Fig.\ \ref{fig5}.

To conclude, we have studied the dynamics of quantum systems in the
presence of stochastic resets and have shown that the steady state
density matrices, averaged over reset probability distribution,
retains off-diagonal terms. We also show that such dynamics for
integrable models leads to family of GGEs/stationary states
characterized by a reset probability $r$. We have discussed
experiments using ultracold bosons which can test our theory.

We note that after our work was completed, Ref.\
[\onlinecite{garrahan1}], which studies the spectral properties of
quantum systems under random resets using a different approach,
appeared on the arXiv.

\vspace{.5cm} \centerline{\bf Acknowledgments} \vspace{.5cm}

S.N.M acknowledges support from ANR grant ANR-17-CE30-0027-01
RaMaTraF and the Indo-French Centre for the promotion of advanced
research (IFCPAR) under Project No. 5604-2.

\vspace{.5cm}

\appendix

\section{Application to single particle quantum mechanics}
\label{qmsec}

In this appendix, we are going to chart out the effect of reset on
two simple single particle quantum mechanical systems. The first
constitutes the evolution of a Gaussian wavepacket under reset while
the second involves that of a coherent state of a simple harmonic
oscillator.

For the first case, we consider a 1D Gaussian wavepacket whose
normalized wavefunction at $t=0$ is given by
\begin{eqnarray}
\psi(x,0) &=& \exp[-x^2/(2\sigma^2)]/\sqrt{2\pi \sigma} \nonumber\\
&=& \frac{\sqrt{\sigma}}{2 \pi} \int dk e^{i k x} e^{-k^2
\sigma^2/2} \label{wpacketeq1}
\end{eqnarray}
where $\sigma$ quantifies the spread of the wavepacket in real
space. For a free particle with $H = k^2/(2m)$, the
wavefunction at any time $t$ is given by
\begin{eqnarray}
\psi(x,t) &=& \frac{\sqrt{\sigma}}{\sqrt{2 \pi (\sigma^2 + i
t/m) }} e^{-\frac{x^2}{2(\sigma^2+ i
t/m)}}\label{wpacketeq2}
\end{eqnarray}
Note that this indicates a ballistic spread of the wavepacket as is
customary in quantum mechanics. Now consider an evolution with reset
characterized by the reset time distribution $p(\tau|t_0)$ given in
Eq.\ (\ref{probeq2}), with the measurement time $t_0\to \infty$.
Then the stationary probability density, characterized by the square
of the absolute value of wavefunction averaged over the reset time
distribution, is given by
\begin{eqnarray}
P(x;r) &=& \int_0^{\infty} d \tau r e^{-r \tau} |\psi(x,\tau)|^2
\label{wpacketeq3}
\end{eqnarray}
For large $x$, the integral in Eq.\ (\ref{wpacketeq3}) can be
evaluated by saddle point method and yields $P(x;r) \simeq \exp[-c
(|x|^2/r)^{1/3}]$, where $c$ is a constant independent of $r$. This
stretched exponential behavior of the tail of the probability
distribution is to be contrasted with its counterpart for a
diffusive classical system for which $P \sim \exp[- c'
|x^2/r|^{1/2}]$ where $c'$ is a constant~[\onlinecite{EM12011}]. The
difference in these two behaviors originates from the ballistic
nature of the spread of the wavepacket in the quantum case.

For the second case, we consider a coherent state for a simple
harmonic oscillator given by
\begin{eqnarray}
|\alpha(t)\rangle &=& e^{-|\alpha|^2/2} \sum_n \frac{\alpha^n
}{{\sqrt n !}} e^{-i \omega_n t} |n\rangle  \label{coheq1}
\end{eqnarray}
where $\omega_n = E_n = \omega_0 (n+1/2)$ denotes frequencies
corresponding to harmonic oscillator energy levels, $\omega_0$ is
the natural oscillator frequency.

Now consider a typical element of the density matrix constructed out
of this coherent state wavefunction. This is given by
\begin{eqnarray}
\rho_{mn}= e^{-|\alpha|^2} \frac{[\alpha^n
(\alpha^*)^m]}{\sqrt{m!n!}} e^{i (m-n) \omega_0 t} \label{coheq2}
\end{eqnarray}
In the absence of any reset, the long time average of any off
diagonal terms vanishes. This leads to the diagonal ensemble.
However, if we now introduce Stochastic reset with $P(r) =r e^{-rt}$
we find finite off-diagonal elements
\begin{eqnarray}
\langle \rho_{mn}\rangle = e^{-|\alpha|^2}\frac{[\alpha^n
(\alpha^*)^m]}{\sqrt{m! n!}} \frac{r(r- i \omega_0
(n-m))}{r^2+(n-m)^2\omega_0^2}  \label{coheq3}
\end{eqnarray}
The presence of such finite off-diagonal elements is manifested in
several physical quantities. For example, the mean position of the
wavepacket, without reset, is given by (assuming real $\alpha$
without loss of generality) $X(t) = \sqrt{2/\omega_0} \alpha \cos(
\omega_0 t)$. Note that the time average of $X$ vanishes signifying
localization of the wavepacket; thus the diagonal ensemble result
corresponds to $X(t \to \infty)=0$. In contrast, the time average of
$X$ with the reset is
\begin{eqnarray}
{\overline X}(r) &=& \int_0^{\infty} dt r e^{-r t} X(t) = \sqrt{
\frac{2 \alpha^2}{\omega_0}} \frac{r^2}{r^2+\omega_0^2}
\label{coheq4}
\end{eqnarray}
The result interpolates between Zeno ($r \to \infty$) and diagonal
ensemble ($r \to 0$) limits as expected. This demonstrates that the
mean position of the coherent state wavepacket can be controlled by
the reset rate $r$.

\section{Derivation of the scaling function
$F(x)$ in Eq.\ (\ref{scaling_function.2})} \label{Fx_appendix}

We start from the exact expression for $n_m(r)$ in Eq.\
(\ref{fchaineq1.5}), where
\begin{equation}
C_k(r)= r \int_0^{\infty} J_k^2(\tau)\, e^{-r\tau}\, d\tau\, .
\label{A.ck1}
\end{equation}
Next we use the exact identity~[\onlinecite{martin2008}]
\begin{equation}
\int_0^{\infty} J_k^2(t)\, e^{-r \tau}\, d\tau= \frac{1}{4\pi}\int_{-\pi}^{\pi} d\theta\,
\frac{e^{i k \theta}}{\sqrt{\frac{r^2}{4}+ \sin^2\left(\frac{\theta}{2}\right)}}\, .
\label{A_iden.1}
\end{equation}
Making the change of variable $k\theta=q$ gives
\begin{equation}
C_k(r)=\frac{r}{4\pi k}\, \int_{-k\pi}^{k\pi} dq \frac{e^{iq}}{\sqrt{\frac{r^2}{4}+
\sin^2\left(\frac{q}{2k}\right)}}\,.
\label{A_iden.2}
\end{equation}
We now take the scaling limit, $r\to 0$, $k\to \infty$, while keeping the product $y=rk$ fixed.
Setting $k=y/r$ with $y$ fixed we get
\begin{equation}
C_{k=y/r}(r)= \frac{r^2}{4\pi\, y}\,\int_{-\pi\,y/r}^{\pi \,y/r} dq \frac{e^{iq}}{\sqrt{\frac{r^2}{4}+
\sin^2\left(\frac{qr}{2y}\right)}}\, .
\label{A_iden.3}
\end{equation}
In the limit $r\to 0$ (with fixed $y$), we can send the limits of integrations to $\pm \infty$
and also expand the sine in the denominator to leading order for small argument. This leads to
\begin{equation}
C_{k=y/r}(r)\approx \frac{r}{2\pi} \int_{-\infty}^{\infty} \frac{dq\, e^{iq}}{\sqrt{q^2+y^2}}\,.
\label{A_iden.4}
\end{equation}
The integral can be recognized as $2\,K_0(|y|)$. Hence, we get the
result in the scaling limit
\begin{equation}
C_k(r) \approx \frac{r}{\pi}\, K_0(r|k|)\, .
\label{A_iden.5}
\end{equation}
Finally, from Eq.\ (\ref{fchaineq1.5}), we get in the scaling limit
(where the sum in Eq.\ (\ref{fchaineq1.5}) can be replaced by an
integral as $r\to 0$)
\begin{eqnarray}
n_m(r) = \sum_{k=m}^{\infty} C_k(r) &\approx& \sum_{k=m}^\infty r\,
K_0(r|k|) \nonumber \\
&\approx & \frac{1}{\pi}\int_{rm}^{\infty} K_0(|y|)\, dy
\label{A_iden.6})
\end{eqnarray}
This gives the result in Eqs.\ (\ref{scaling_function.1}) and
(\ref{scaling_function.2}).


\begin{thebibliography}{99}

\bibitem{rev1}J. Dziarmaga, Adv. Phys. {\bf 59}, 1063 (2010).

\bibitem{rev2} A. Polkovnikov, K. Sengupta, A. Silva, and M. Vengalattore,
Rev. Mod. Phys. {\bf 83}, 863 (2011).

\bibitem{rev3} A. Dutta, U. Divakaran, D. Sen, B.K. Chakrabarti, T.F. Rosenbaum, G. Aeppli,
arXiv:1012.0653 (unpublished); S. Mondal, D. Sen and K. Sengupta,
Lect. Notes Phys. {\bf 21}, 802 (2010).

\bibitem{rev4} I. Bloch, J. Dalibard, and W. Zwerger, Rev. Mod. Phys. {\bf 80}, 885
(2008).

\bibitem{rev5} L. D'Alessio, Y. Kafri, A. Polkovnikov and M. Rigol,
Adv. Phys. {\bf 65}, 239 (2016).

\bibitem{subir0} K. Sengupta, S. Powell, and S. Sachdev, \pra {\bf 69}, 053616 (2004)

\bibitem{pascal1} P. Calabrese and J. Cardy, Phys. Rev. Lett. {\bf 96}, 136801
(2006); {\it ibid}, J. Stat. Mech. {\bf 0706} P06008 (2007).

\bibitem{kibble1} T.W.B. Kibble, J. Phys. A: Math. Gen. {\bf 9}, 1387
(1976); {\it ibid}, Phys. Rep. {\bf 67}, 183 (1980)

\bibitem{zureck1} W.H. Zurek, Nature (London) {\bf 317}, 505 (1985); {\it
ibid} Phys. Rep. {\bf 276}, 177 (1996); B. Damski, Phys. Rev. Lett.
{\bf 95}, 035701 (2005); W.H. Zurek, U. Dorner, and P. Zoller, Phys.
Rev. Lett. {\bf 95}, 105701 (2005)

\bibitem{anatoly1} A. Polkovnikov, Phys. Rev. B {\bf 72}, 161201(R)
(2005); R. Barankov, A. Polkovnikov, Phys. Rev. Lett. {\bf 101},
076801 (2008); A. Chandran, A. Erez, S.S. Gubser, S.L. Sondhi, Phys.
Rev. B {\bf 86}, 064304 (2012).

\bibitem{ks1} K. Sengupta, S. Mondal, and D. Sen, Phys. Rev. Lett. {\bf 100}, 077204
(2008); S. Mondal, D. Sen and K. Sengupta, Phys. Rev. B {\bf 78}
(2008) 045101.

\bibitem{ks2} D. Sen, S. Mondal, K. Sengupta, Phys. Rev. Lett. {\bf 101}, 016806
(2008);

\bibitem{ks3} J.D. Sau, K. Sengupta, Phys. Rev. B {\bf 90}, 104306 (2014);
U. Divakaran, K. Sengupta, Phys. Rev. B 90, 184303 (2014).

\bibitem{degrandi1} C. De Grandi and A. Polkovnikov,
in {\it Quantum Quenching, Annealing, and Computation}, edited by A.
K. Chandra, A. Das, and B. K. Chakrabarti, Lecture Notes in Physics,
{\bf 802}, 75 (Springer, Heidelberg, 2010).

\bibitem{sumit1} S.R. Das, D.A. Galante and R.C. Myers, Phys. Rev. Lett. {\bf 112} 171601
(2014); {\it ibid}, JHEP {\bf 02} 167 (2015); D. Das, S. R. Das, D.
A. Galante, R. C. Myers, and K. Sengupta, JHEP {\bf 11}, 157 (2017).

\bibitem{mrigol1} M. Rigol, V. Dunjko and M. Olshanii, Nature {\bf 452} 854 (2008).

\bibitem{mrigol2} J.M. Deutsch, Phys. Rev. A {\bf 43} 2046 (1991).

\bibitem{srednicci1} M. Srednicki, Phys. Rev. E {\bf 50}, 888
(1994); {\it ibid} J. Phys. A {\bf 32}, 1163 (1999).

\bibitem{arnabd1}A. Lazarides, A. Das, and R. Moessner, Phys. Rev. Lett. {\bf 112},
150401 (2014); {\it ibid}, Phys. Rev. E {\bf 90}, 012110 (2014); P.
Ponte, A. Chandran, Z. Papic, and D. A. Abanin,  Ann. Phys.
(Amsterdam) {\bf 353}, 196 (2014); L. D'Alessio and M. Rigol, Phys.
Rev. X {\bf 4}, 041048 (2014)

\bibitem{asen0} S. Nandy, A. Sen, and D. Sen, Phys. Rev. X {\bf 7},
031034 (2017).

\bibitem{EM12011} M. R. Evans and S. N. Majumdar, \prl {\bf 106}, 160601 (2011).

\bibitem{EM22011} M. R. Evans and S. N. Majumdar, J. Phys. A: Math. Theor. {\bf 44}, 435001 (2011).

\bibitem{EM32014} M. R. Evans and S. N. Majumdar, J. Phys. A: Math. Theor. {\bf 47}, 455004 (2014).

\bibitem{GMS2014} S. Gupta, S. N. Majumdar, and G. Schehr, Phys. Rev. Lett.
112, 220601 (2014).

\bibitem{DHP2014} X. Durang, M. Henkel, and H. Park, J. Phys. A: Math. Theor.
{\bf 47}, 045002 (2014).

\bibitem{MSS2015} S. N. Majumdar, S. Sabhapandit, and G. Schehr, Phys. Rev. E
{\bf 91}, 052131 (2015).

\bibitem{MSS22015} S. N. Majumdar, S. Sabhapandit, and G. Schehr, Phys. Rev. E
{\bf 92}, 052126 (2015).

\bibitem{Pal2015} A. Pal, Phys. Rev. E {\bf 91}, 012113 (2015).

\bibitem{EM2016} S. Eule and J. J. Metzger, New J. Phys. {\bf 18}, 033006
(2016).


\bibitem{NG2016} A. Nagar and S. Gupta, Phys. Rev. E {\bf  93}, 060102 (2016).

\bibitem{BEM2016} D. Boyer, M. R. Evans, and S. N. Majumdar, J. Stat. Mech. P023208 (2017).

\bibitem{PKE2016} A. Pal, A. Kundu, and M. R. Evans, J. Phys. A. Math. Theor.
{\bf 49}, 225001 (2016).

\bibitem{FBGM2017} A. Falcon-Cortes, D. Boyer, L. Giuggioli, and
S. N. Majumdar, Phys. Rev. Lett. {\bf 119}, 140603 (2017).

\bibitem{MT2017} C. Maes and T. Thiery, J. Phys. A. Math. Theor. {\bf 50}, 415001 (2017).

\bibitem{RG2017} E. Roldan and S. Gupta, Phys. Rev. E {\bf 96}, 022130 (2017).


\bibitem{Rednerbook} S. Redner, {\it A Guide to First-Passage
Processes}, (Cambridge University Press, Cambridge, 2001).

\bibitem{BMSreview} A. J. Bray,  S. N. Majumdar, and G. Schehr,
Adv. in Phys. {\bf 62}, 225 (2013).

\bibitem{EMM2013} M. R. Evans, S. N. Majumdar, and K. Mallick, J. Phys. A: Math.
Theor. {\bf 46}, 185001 (2013).

\bibitem{WEM2013} J. Whitehouse, M. R. Evans, and S. N. Majumdar, Phys. Rev. E
{\bf 87}, 022118 (2013).

\bibitem{MV2013} M. Montero and J. Villarroel, Phys. Rev. E
{\bf 87}, 012116 (2013).

\bibitem{KMSS2014}  L. Kusmierz, S. N. Majumdar, S. Sabhapandit, and G. Schehr,
Phys. Rev. Lett. {\bf 113}, 220602 (2014).

\bibitem{RUK2014} S. Reuveni, M. Urbach, and J. Klafter, Proc. Natl. Acad. Sci.
USA {\bf 111}, 4391 (2014).

\bibitem{CS2015} C. Christou and A. Schadschneider,  J. Phys. A: Math. Theor.
{\bf 48}, 285003 (2015).

\bibitem{R2016} S. Reuveni, Phys. Rev. Lett. {\bf 116}, 170601 (2016).

\bibitem{BBS2016} U. Bhat, C. De Bacco, and S. Redner, J. Stat. Mech. P083401 (2016).

\bibitem{MPV2017} M. Montero, M. Maso-Puigdellosas, and J. Villarroel, Eur. Phys. J. B {\bf 90}, 176 (2017).

\bibitem{PR2017} A. Pal and S. Reuveni, Phys. Rev. Lett. {\bf 18}, 030603 (2017).

\bibitem{MST2015} J. M. Meylahn, S. Sabhapandit, and
H. Touchette, Phys. Rev. E {\bf 92}, 062148 (2015)

\bibitem{HT2017} R. J. Harris and H. Touchette, J. Phys. A: Math. Theor.
{\bf 50},  10LT01 (2017).

\bibitem{HMMT2018} F. den Hollander, S.N. Majumdar,
J. M. Meylahn, and H. Touchette, arXiv: 1801.09909

\bibitem{FGS2016} J Fuchs, S Goldt, and U Seifert, EuroPhys. Lett. {\bf 113}, 6 (2016).

\bibitem{PR22017} A. Pal and S. Rahav, Phys. Rev. E {\bf 96}, 062135 (2017).

\bibitem{sto1} T. Albash, D.A. Lidar, M. Marvian, and P. Zanardi,
Phys. Rev. A 88, 023146 (2013); A.E. Rastegin, J. Stat. Mech. P06016
(2013); D. Kafri and S. Deffner, Phys. Rev. {\bf A} 86, 044302
(2012); M. Campisi, J. Pekola, and R. Fazio, New J. Phys. {\bf 19},
053027 (2017); S. Gherardini, L. Buffoni, M. M. Muller,, F. Caruso,
M. Campisi, A. Trombettoni, and S. Ruffo, arXiv:1805.00773
(unpublished).

\bibitem{dhar1} S. Dhar, S. Dasgupta, A. Dhar, and D. Sen, Phys. Rev. A {\bf 91},
062115 (2015).

\bibitem{eli1} F. Thiel. E. Barkai, and D. A. Kessler, \prl {\bf 120}, 040502
(2018); H. Friedman, D. A. Kessler, and E. Barkai, \pre {\bf 95},
032141 (2017).

\bibitem{zenoref} C. B. Chiu, E. C. G. Sudershan, and B. Misra, Phys. Rev. D {\bf 16},
520 (1977); W. M. Itano, D. J. Heinzen, J. J. Bollinger, and D. J. Wineland
Phys. Rev. A {\bf 41}, 2295 (1990); D. Home and M. A. B Whittaker,
Ann. Phys. {\bf 258}, 237 (1997).

\bibitem{subir1} S. Sachdev, K. Sengupta, and S. M. Girvin, \prb {\bf
66} 075128 (2002); S. Pielawa, T. Kitagawa, E. Berg, and S. Sachdev,
Phys. Rev. B {\bf 83}, 205135 (2011); C. P. Rubbo, S. R. Manmana, B.
M. Peden, M. J. Holland, and A. M. Rey, Phys. Rev. A {\bf 84},
033638 (2011).

\bibitem{greiner1} W. S. Bakr, A. Peng, M. E. Tai, R. Ma, J. I. Gillen, and S. F�ollen,
Science {\bf 329}, 547 (2010).

\bibitem{greiner2} J. Simon, W. Bakr, R. Ma, M. E. Tai, P. Preiss, and M. Greiner,
Nature (London) {\bf 472}, 307 (2011).

\bibitem{antal1} T. Antal, Z. Racz, A. Rakos, and G.~M. Schutz, Phys. Rev. E {\bf 59}, 4912 (1999).

\bibitem{antal2} T. Antal, P.~L. Krapivsky, and A. Rakos, Phys. Rev. E {\bf 78}, 06115 (2008).

\bibitem{eisler1} V. Eisler and Z. Racz, \prl {\bf 110}, 060602
(2013); V. Hunyadi, Z. Racz, and L. Sasvari, \pre {\bf 69}, 066103
(2004).

\bibitem{subirbook1}S. Sachdev, {\it Quantum Phase Transitions}, (Cambridge
University Press, Cambridge, 1999).

\bibitem{kit1} A. Kitaev, Ann. Phys. (N.Y.) {\bf 321}, 2 (2006).

\bibitem{netorev} A. H. Castro Neto, F. Guinea, N. M. R. Peres, K. S. Novoselov, and A. K.
Geim, Rev. Mod. Phys. {\bf 81}, 109 (2009).

\bibitem{topoinrev} M. Z. Hasan and C. L. Kane, Rev. Mod. Phys. {\bf 82}, 3045 (2010).

\bibitem{floquetrev1} A. Eckardt Rev. Mod. Phys. {\bf 89}, 011004
(2017).

\bibitem{asen1} A. Sen, S. Nandy, and K. Sengupta, \prb {\bf 94}, 214301
(2016).

\bibitem{dynamicsdipole} K. Sengupta, S. Powell, and S. Sachdev,
\pra {\bf 69}, 053616 (2004); M. Kolodrubetz, D. Pekker, B. K.
Clark, and K. Sengupta, \prb {\bf 85}, 100505(R) (2012); U.
Divakaran and K. Sengupta, \prb {\bf 90}, 184303 (2014); R. Ghosh,
A. Sen, and K. Sengupta, \prb {\bf 97}, 014309 (2018).

\bibitem{mukund1} Y. S. Patil, S. Chakram, and M. Vengalattore, \prl {\bf 115}, 140402
(2015); Y. S. Patil, S. Chakram, L. M. Aycock, and M. Vengalattore,
\pra {\bf 90}, 033422 (2014).

\bibitem{martin2008} P. A. Martin, J. Phys. A. Math. Theor. {\bf 41}, 015207 (2008).

\bibitem{garrahan1} D. C. Rose, H. Touchette, I. Lesanovsky, and J. P. Garrahan,
arXiv:1806.01298 (unpublished).


\end{thebibliography}
\end{document}